\documentclass[10pt]{revtex4}
\usepackage{graphicx}
\usepackage{bm}
\usepackage{comment} 
\usepackage{amssymb}
\usepackage{color}
\usepackage{amsbsy}      
\usepackage{amsmath}
\usepackage{lineno}
\usepackage{xfrac}
\usepackage{setspace}
\def\dd{\mbox{d}}

\begin{document}

\title{Age-structured social interactions enhance radicalization}

\author{Yao-Li Chuang}
 \affiliation{UCLA Department of Biomathematics and CSUN Department of
   Mathematics} \

\author{Tom Chou}
 \affiliation{UCLA Department of Biomathematics and Department of
   Mathematics} \

\author{Maria R. D'Orsogna}
 \email{dorsogna@csun.edu}
 \affiliation{UCLA Department of Biomathematics and CSUN Department of
   Mathematics~} \ \date{\today}




\begin{abstract}
\noindent
Disaffected youth are among the most susceptible in espousing and
acting on extremist ideals, as confirmed by demographic studies. To
study age-dependent radicalization we introduce a three-stage model
where individuals progress through non-radical, activist, and radical
states, while also aging.  Transitions between stages are modeled as
age-dependent interactions that are maximized for individuals of the
same age and that are enhanced at early adulthood. For comparison, we
also derive the age-independent formulation corresponding to the full
age-dependent model.  We find that age-dependence leads to more
complex dynamics, enhancing radicalization in certain parameter
regimes. We also observe waves of radical behavior ebbing and flowing
over generational cycles, realizing well known paradigms in political
science.  While government intervention is most effective when the
appropriate ages are targeted, deciding whether preventive or
corrective action is preferable depends on the aggressiveness of the
radicalization process.
\end{abstract}
\keywords{radicalization | age dependence | population dynamics
        | social interaction | mathematical model | differential equation}

\maketitle

\section{Introduction}

Understanding why and how individuals adopt extremist views has become
a prime focus for governments and society. The process commonly known
as ``radicalization'' is gradual, unfolds through several phases, and
may be strongly influenced by social, economic and cultural factors
\cite{SPR91,SIL07,MCA08,SIL08}.  The making of a radical often starts
from a state of discontent and feelings of alienation.  Mainstream
political, social, or religious tenets are gradually rejected and
replaced by alternate and increasingly extreme ideologies
\cite{SIL03,HOR08,KIN11, KOE14}.  As new ideals become entrenched, a
phase of intolerance towards the beliefs, identities, and lifestyles
of ``others" ensues, accompanied by a proselytizing push \cite{VKN07}.
Finally, radicalization may culminate with the execution of violent,
self-destructive acts aimed at spreading terror and damage
\cite{CRE11,MEE15}.  Although the boundaries between the above
described stages are somewhat fluid, we may succinctly describe
radicalization as a sequence of pre-radicalization,
self-identification, indoctrination, commitment and jihadization
steps; each being marked by an increasing level of fanaticism
\cite{BOR03,WIK04,MOG05,PRE07,SAG08,KLA16}.

Radical tendencies can arise at any age and factors that are
traditionally associated with desistance from deviant behavior, such
as marriage, career and education, are not always sufficient
deterrents against extremism \cite{SAG04, KIN11}. Disenfranchised
young adults, however, are particularly vulnerable to indoctrination
and radicalization, especially when their formative years are spent
without purpose, education or positive role models \cite{LOF65,
  MEE15}.  Joining extremist groups can provide a sense of belonging
and a haven where ideals are shared with like minded individuals
\cite{SEU99,SAG04, HAR05, BAK06, LOZ11}.  Today, this is sometimes the case for
marginalized immigrant or second-generation youth in western cities,
for those growing up in refugee camps, and for right-wing extremists
\cite{DOO13,KOE14}. Lacking prospects or motivation, an aversion for the
majority and a desire to undermine authority gradually takes root with
radicalism providing a sense of purpose and community.

Several mathematical models have been proposed to describe
radicalization, both as a general phenomenon \cite{CAS03, MAR12,
  CAM13,SHO17}, and as applied to actual situations, such as the emergence
of separatist movements in the Basque country \cite{EHR13}, or of
right-wing groups in Germany \cite{DEU15}.  These models include
temporal and in some cases spatial patterns of populations evolving
through stages of increasing fanaticism
\cite{SAG04,BAK06,LOZ11,AHM13}.  In some instances field data was used
to recapitulate violent incidents that were later used for parameter
estimation \cite{SAN08}.  Despite the many useful insights provided,
none of these studies include age-sensitive responses to propaganda,
emulation of peers and societal pressure.  As described above,
radicalization is highly age-dependent and developing age-structured
models may shed new light on the mechanisms that lead to the
establishment of radical groups and help identify optimal intervention
strategies.

In this paper we propose an age-structured model for radicalization
where a sequence of stages marked by increasing fanaticism is coupled
with age-differentiated interactions.  Age dependence intensifies
interactions among peers, enhancing the progression towards extremist
behavior.  Our results suggest that conventional age-dependent
population models may oversimplify the complexity of social
interactions.  In some parameter regimes, we find that age-structure
leads to enhanced radicalization, and that for societies that are
highly prone to irreversible radicalization, generational cycles of
extremism may arise, realizing a well known paradigm in political
science of alternating decades of ebbing and flowing radicalization
\cite{RAP02}. We also study the effects of government intervention and
find that strategically focusing resources on specific age ranges may
lead to optimal results.

\section{The model}

\begin{figure} [h]
  \begin{center}
      \includegraphics[width=3.3in]{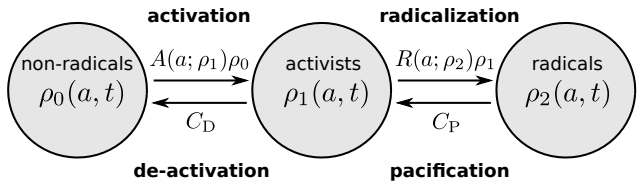}
\end{center}
  \caption{Illustration of the population exchange among non-radicals
    $\rho_0$, activists $\rho_1$, and radicals $\rho_2$. Activation
    and radicalization are assumed to occur through social
    interactions between converting and converted population
    fractions. De-activation and pacification are proportional to the
    converted population fractions.
         \label{FIG:DIAGRAM} }
\end{figure}

\noindent
Several classifications have been introduced in the literature to
describe the number of stages between pre-radicalization and full
fledged extremism
\cite{BOR03,WIK04,MOG05,PRE07,SIL07,MCA08,SAG08,KIN11}.  In this work,
for simplicity, we consider three distinct stages of fanaticism
labeled as $i=0,1,2$.  The first $i=0$ stage is that of
pre-radicalization, where individuals do not respond to extreme
ideologies and are referred to as ``non-radicals.'' Upon being exposed
to radical ideas, non-radicals may choose to espouse them, become
``activists,'' and recruit others to their newly found ideology.  This
intermediate $i=1$ state corresponds to self-identification and early
indoctrination.  Finally, $i=2$ is the last stage whereby activists
turn into ``radicals'' who embrace violence to further their cause.
The transitions among the three stages are illustrated in
Fig.~\ref{FIG:DIAGRAM}.

To introduce an age-structure we define non-radical, activist and
radical population densities $\rho_i (t, a)$ of age $a$ at time $t$.
Transitions from the non-radical $i=0$ to the activist $i=1$ pool for
individuals of age $a$ are mediated by the activation rate $A(a;
\rho_1)$ which we assume depends on $a$ and on the age-structure of
the influencing pool $\rho_1$.  Similarly, transitions from the
activist $i=1$ to the radical $i=2$ population at age $a$ are driven
by the radicalization rate $R (a; \rho_2)$. We finally include a
de-activation rate $C_D$ that leads individuals to regress from $i=1$
to $i=0$ and a pacifying rate $C_{\rm P}$ for individuals to revert
from $i=2$ to $i=1$.  Both $C_{\rm D}$ and $C_{\rm P}$ could be age
structured, but we keep them uniform for simplicity. 
Numerical studies with monotonically increasing $C_{\rm D}(a)$ and $C_{\rm P}(a)$ 
do not yield qualitatively different results and are not shown here.
We write our model as 

\begin{eqnarray}
 \label{EQ:RHO0} 
   \frac{\partial \rho_0}{\partial t} + \frac{\partial \rho_0 }{\partial a} &=& 
   - A (a; \rho_1) \rho_0 + C_{\rm D} \rho_1,  \\
   \frac{\partial \rho_1} {\partial t} + \frac{\partial \rho_1 }{\partial a} &=& A (a; \rho_1) \rho_0 
    - \left[ C_{\rm D} + R(a; \rho_2) \right] \rho_1 +
  \nonumber \\
    \label{EQ:RHO1}  
   &&  C_{\rm P} \rho_2.
\\
  \frac{\partial \rho_2 }{\partial t} + \frac{\partial \rho_2 }{\partial a} &=&  
   \label{EQ:RHO2} 
   R (a; \rho_2) \rho_1  - C_{\rm P} \rho_2. \\
   \nonumber
\end{eqnarray}

\noindent
Eqs.\,\ref{EQ:RHO0} -- \ref{EQ:RHO2} are of the McKendrick-von
Foerster type \cite{MCK26,LES45,LES48,TRU65,KEY97,CHO16}, where the
left-hand side is the total time derivative $\sfrac{\rm d}{{\rm d}t} =
\sfrac{\partial}{\partial t} + \left( \sfrac{\partial a}{\partial t}
\right) \sfrac{\partial}{\partial a}$. Provided that age and time are
measured in the same unit, $\sfrac{\partial a}{\partial t} = 1$, and
the $\sfrac{\partial}{\partial a}$ term is associated to ``aging''.
The transition rates $A(a; \rho_1)$ and $R (a; \rho_2)$ on the
right-hand side depend explicitly on age and are defined as

\begin{eqnarray}
A (a; \rho_1) &=& C_{\rm A} \int_{a_0}^{a_1}   
   {\cal K} (a,a'; \alpha_A, \sigma_A) \, 
   \rho_1 (t, a^\prime) \dd a^\prime,
   \label{EQ:ACTIVATION_RATE} \\
R (a; \rho_2) &=&  C_{\rm R} \int_{a_0}^{a_1} 
   {\cal K} (a,a'; \alpha_R, \sigma_R) \, 
   \rho_2 (t, a^\prime) \dd a^\prime.
   \label{EQ:RADICALIZATION_RATE}
\end{eqnarray}

\noindent
Eqs.\,\ref{EQ:ACTIVATION_RATE} and \ref{EQ:RADICALIZATION_RATE}
indicate that all individuals between ages $a_0$ and $a_1$ are able to
influence those at age $a$ via the ``interaction kernels'' ${\cal
  K}(a,a', \alpha_j, \sigma_j)$ that we model as

\begin{eqnarray}
\label{kernels}
      {\cal K}(a,a'; \alpha_j, \sigma_j) =
      \frac{\displaystyle{\exp \left[ - \frac{( a - \alpha_j )^2 +
                (a - a^\prime)^2 }{2 \sigma_j^2} \right]}}
   {\displaystyle{\int_{-\bar{a}}^{\bar{a}} \exp \left[ -\frac{s^2}{\sigma_j^2} \right] \dd s}}
\end{eqnarray}

\noindent
for $j = {\rm A, R}$, where $\bar{a} \equiv \vert a_1 - a_0 \vert$.
The kernels specify how an individual at age $a$ is influenced by
another at age $a^\prime$.  Eq.\,\ref{kernels} is defined so that if
$\rho_1 (t, a^\prime)$ in Eq.\,\ref{EQ:ACTIVATION_RATE} is
age-independent, then $A (a; \rho_1) \to C_{\rm A}$ and similarly if
$\rho_2 (t, a^\prime)$ is age-independent, then $R (a; \rho_2) \to
C_{\rm R}$.  The kernels are maximized when $a = \alpha_j$ and when $a
= a^\prime$. The former condition expresses that individuals are most
susceptible to activation and radicalization when they are at certain
``target'' ages $\alpha_j$ ($j={\rm A,R}$). The latter condition
arises from ``peer-to-peer'' interactions, whereby individuals are
most influenced by those of similar age.
Eq.\,\ref{EQ:ACTIVATION_RATE} decreases as $a$ shifts away from
$\alpha_j$ and when the age gap $\vert a - a^\prime \vert$
increases. How quickly these kernels decline is specified by
$\sigma_j$ ($j={\rm A,R}$), the spread of the interaction kernel. If
$\sigma_j$ is large, the influence exerted by individuals at age
$a^\prime$ on those at age $a$ may persist for large age differences
$\vert a^\prime - a \vert$, and for large $\vert a - \alpha_j \vert$.
Conversely, for small $\sigma_j$ the kernels will be appreciable only
for $a \sim a^\prime$ and $a \sim \alpha_j$.  Furthermore, with
Eq.\,\ref{EQ:ACTIVATION_RATE} we assume that activation ($i = 0$ to
$i=1$) occurs through age-dependent social interactions between the
$\rho_0$ non-radical and the activist $\rho_1$ populations; a similar
construct holds for radicalization ($i=1$ to $i=2$) in
Eq.\,\ref{EQ:RADICALIZATION_RATE} through which the $\rho_1$ activists
interact with the $\rho_2$ radicals.  We will study the model defined
by Eqs.\,\ref{EQ:RHO0}--\ref{kernels} for a typical age span of
$\left[ a_0, a_1 \right]$, so that $\bar{a} \equiv a_1 - a_0$ in the
denominator of Eq.\,\ref{kernels}.  The latter is introduced to
guarantee that upon integration over $[a_0, a_1]$ the kernels are
independent of $\sigma_{\rm j}$. This will be useful when deriving the
age-independent version of Eqs.\,\ref{EQ:RHO0}--\ref{kernels}, since
the denominator will yield age-independent transition rates that
depend only on the $C_j$ amplitudes and not on $\sigma_j$.  It is
important to note that the presence of the denominator implies that
the total area under the kernel is unity; kernels with a higher
$\sigma_j$ will be wider but shallower than kernels with a lower
$\sigma_j$.

The chosen $[a_0, a_1]$ boundaries specify the range for effective
age-based interactions: individuals of age $a < a_0$ are too young to
influence or be influenced by an ideology or by their peers, those
with age $a > a_1$ may be too old or entrenched for change.  The age
boundaries also imply that Eqs.\,\ref{EQ:RHO0}-\ref{kernels} represent
radicalization across a single generation, beginning at $a_0$ and
ending at $a_1$.  For simplicity we assume that within $[a_0, a_1]$
death is negligible. To complete our model we must include boundary
conditions which are chosen at $a=a_0$ as

\begin{eqnarray}
  \rho_0 (t, a_0) & = & \sum_{i=0,1,2} \rho_i (t, a_1), \label{EQ:BIRTH1} \\
  \rho_1 (t, a_0) & = & \rho_2 (t, a_0) = 0. \label{EQ:BIRTH2} 
\end{eqnarray}

%
\begin{figure*}[t]
  \begin{center}
      \includegraphics[width=6.9in]{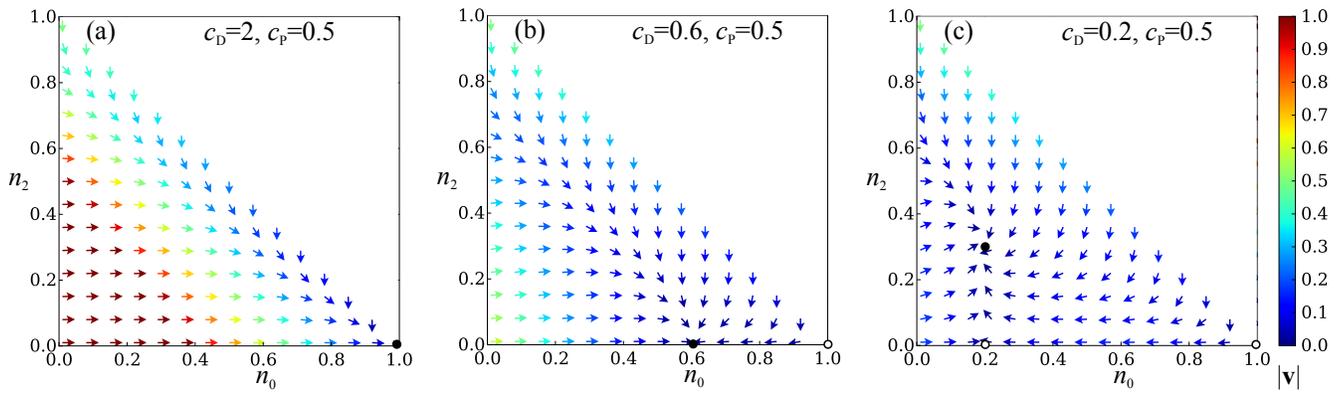}
\end{center}
\caption{Reversible radicalization phase portraits derived from
  Eqs.\,\ref{EQ:NONACTIVIST1} and \ref{EQ:RADICAL1} for three values
  of $c_{\rm D}$ and fixed $c_{\rm P} = 0.5$, $c_{\rm A} = c_{\rm R} =
  1$.  The magnitude of the phase flow $\mathbf{v} \equiv \left( {\rm
    d} n_0 / {\rm d} t, {\rm d} n_2 / {\rm d} t \right)$ is
  color-coded.  (a) For $c_{\rm D} / c_{\rm A} > 1$ all trajectories
  converge to the stable fixed point at $(n_0^*, n_2^*) = (1, 0)$
  denoted by a solid circle. (b) For $ 1 - c_{\rm P}/c_{\rm R} <
  c_{\rm D}/c_{\rm A} \leq 1$ the stable fixed point is $(n_0^*,
  n_2^*) = (c_{\rm D}/c_{\rm A}, 0)$ while $(1, 0)$ turns unstable and
  is denoted by an open circle.  (c) Finally, for $c_{\rm D}/ c_{\rm
    A} \leq 1 - c_{\rm P}/c_{\rm R}$ the sole stable solution is
  $(n_0^*, n_2^*) = \left( c_{\rm D}/c_{\rm A}, 1 - c_{\rm D}/c_{\rm
    A} - c_{\rm P}/c_{\rm R} \right)$, while the other two are
  unstable.  The regions above $n_0+n_2 = 1$ are unphysical as $0 \leq
  n_i \leq 1$ for $i=0,1,2$ by construction.
  \label{FIG:NO_AGE}}
\end{figure*}

\noindent
The above conditions imply that the total population exiting from age
$a=a_1$, including radicals, activists and non-radicals, will re-enter
the system at age $a_0$ as non-radical.  The model is now complete as
Eqs.~\ref{EQ:RHO0}-\ref{EQ:RHO2} are advection-reaction equations that
require boundary conditions only on one side of the domain. Our basic
assumption is that at each $i=0,1,2$ stage the population is large and
stochastic effects can be neglected, justifying the use of
deterministic, mean-field McKendrick-von Foerster type equations
\cite{MCK26,LES45,LES48,TRU65,KEY97,CHO16}.  We also neglect birth and
death events for simplicity, assuming a constant total population and
focusing on how this constant population is distributed across ages
and radicalization stages in response to different social interaction
mechanisms.

Finally, we introduce the size of the population at stage $i$ between
ages $a_1$ and $a_2$ at time $t$ as $N_i (t; [a_1, a_2] ) \equiv
\int_{a_1}^{a_2} \rho_i (t, a) \dd a $, a quantity that will be
invoked in the next section.  Unless otherwise stated we will consider
initial conditions such that $\rho_0(0, a) \gg \rho_1(0, a) \simeq
\rho_2(0,a) > 0$, to mimic a society where the majority of the
population is initially non-radical.  Note that because of the
structure of Eqs.\,\ref{EQ:RHO0}-\ref{EQ:RHO2}, $\rho_1(0,a)$ and
$\rho_2(0,a)$ must be non-zero to induce activation and
radicalization.

\section{Results}

\noindent
We begin by deriving the age-independent equations corresponding to
Eqs.\,\ref{EQ:RHO0}--\ref{EQ:RHO2}. These will give us a benchmark
against which the full, age-structured model will be later compared
and contrasted to.  What we will obtain is an ODE system where kernels
and transition rates are condensed in an age-uniform
quantity. Comparing results from both formulations will be useful to
understand what features we fail to capture when neglecting
age-structure.

\subsection{Age-independent model and stationary solutions}
\label{AI}

\noindent
To construct the age-independent version of our model we define the
fractional populations $n_i (t)$ as

\begin{eqnarray}
\label{nidef}
n_i(t) \equiv \frac {\int_{a_0}^{a_1} \rho_i (t, a) \dd a}
{\displaystyle{\sum_{i = 0,1,2} N_i (t; [a_0,a_1])}}
\end{eqnarray}

\noindent
where the denominator $N \equiv \sum_{i = 0,1,2} N_i (t; [a_0,a_1])$
represents the entire population between ages $a_0$ and $a_1$. The
resulting $N$ is time independent due to population conservation. By
construction $0 \leq n_i(t) \leq 1$ for all $i$ and $\sum_{i=0,1,2}
n_i(t) = 1$.  We now let $\sigma_j \rightarrow \infty$ in
Eq.\,\ref{kernels} for $j = {\rm A, R}$. The interaction kernels
become constants and the age-independent transition rates can be
rewritten as $A(\rho_1) = C_{\rm A} \int_{a_0}^{a_1} \rho_1 (t, a) \dd
a / (2 \bar{a}) = N C_{\rm A} n_1(t)/ (2 \bar a)$, and $R(\rho_1) =
C_{\rm R} \int_{a_0}^{a_1} \rho_2 (t, a) \dd a / (2 \bar{a}) = N
C_{\rm R} n_2(t)/ (2 \bar a)$.  Finally, integrating
Eqs.\,\ref{EQ:RHO0}--\ref{EQ:RHO2} over the ages between $a_0$ and
$a_1$ and dividing the results by $N$ we find that the interactions
factor as

\begin{eqnarray}
  \frac{\dd n_0}{\dd t} & = & - c_{\rm A} n_0 n_1 + c_{\rm D}
  n_1, \label{EQ:NONACTIVIST} \\ \frac{\dd n_1}{\dd t} & = & c_{\rm A}
  n_0 n_1 - c_{\rm D} n_1 - c_{\rm R} n_1 n_2 + c_{\rm P}
  n_2, \label{EQ:ACTIVIST} \\ \frac{\dd n_2}{\dd t} & = & c_{\rm R}
  n_1 n_2 - c_{\rm P} n_2, \label{EQ:RADICAL}
\end{eqnarray}

\noindent
where $c_{\rm A} \equiv N C_{\rm A} /(2 \bar{a})$, $c_{\rm R} \equiv N
C_{\rm R} / (2 \bar{a})$, $c_{\rm D} \equiv C_{\rm D}$, and $c_{\rm P}
\equiv C_{\rm P}$. Note that there is no equivalent to the boundary
conditions Eqs.\,\ref{EQ:BIRTH1} and \ref{EQ:BIRTH2} since we
integrated out
age-dependence. Eqs.\,\ref{EQ:NONACTIVIST}--\ref{EQ:RADICAL} form a
degenerate system, which, by substituting $n_1 = 1 - n_0 - n_2$,
reduces to

\begin{eqnarray}
  \frac{\dd n_0}{\dd t} & = & \left( 1 - n_0 - n_2 \right) \left(
  c_{\rm D} - c_{\rm A} n_0 \right), \label{EQ:NONACTIVIST1}
  \\ \frac{\dd n_2}{\dd t} & = & n_2 \left[ c_{\rm R} \left( 1 - n_0 -
    n_2 \right) - c_{\rm P} \right]. \label{EQ:RADICAL1}
\end{eqnarray}

\begin{figure*}[t]
  \begin{center}
      \includegraphics[width=6.9in]{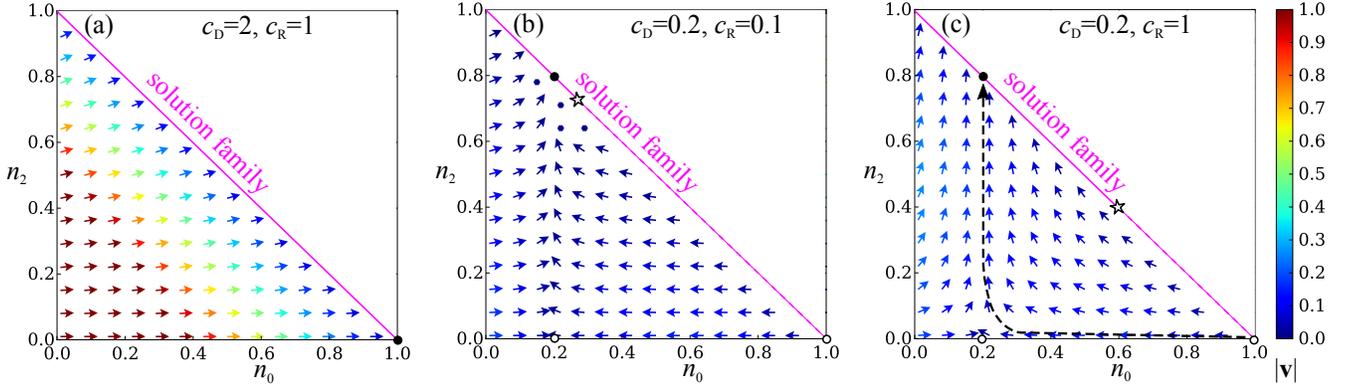}
\end{center}
  \caption{Irreversible radicalization phase portraits derived from
    Eqs.\,\ref{EQ:NONACTIVIST1} and \ref{EQ:RADICAL1} for three sets
    of $(c_{\rm D}, c_{\rm R})$ and fixed $c_{\rm P} =0$, $c_{\rm A} =
    1$.  A family of solutions $n_0^* + n_2^* = 1$ (depicted in
    magenta) arises in addition to the three fixed points. (a) For
    $c_{\rm D} / c_{\rm A}> 1$, the solution family and so the fixed
    point $(1,0)$ are marginally stable. (b)--(c) For $c_{\rm D} /
    c_{\rm A} \leq 1$, the portion of the solution family that
    satisfies $n_0^* > (c_{\rm D} + c_{\rm R})/(c_{\rm A} + c_{\rm
      R})$ becomes unstable (dashed magenta line), along with the
    fixed points $(1,0)$ and $\left( c_{\rm D}/c_{\rm A}, 0 \right)$.
    The remainder of the solution family and the fixed point $\left(
    c_{\rm D}/c_{\rm A}, 1 - c_{\rm D}/c_{\rm A} \right)$ are
    marginally stable (solid magenta line).  In panel (b) where
    $c_{\rm R} = 1.0$ the stable portion of the solution family is
    larger than in panel (c) where $c_{\rm R} = 0.1$.  Phase flows
    steer trajectories towards the solution family, avoiding the
    stable fixed point. A trajectory starting near $(1, 0)$ and ending
    at $(c_{\rm D}/c_{\rm A}, 1 -c_{\rm D}/c_{\rm A})$ is depicted in
    panel (c) as a black dashed curve. Other trajectories may end
    anywhere along the stable portion of the solution family.
               \label{FIG:NO_AGE1} }
\end{figure*}

\noindent
The model defined by Eqs.\,\ref{EQ:NONACTIVIST1} and \ref{EQ:RADICAL1}
is similar to previous multi-compartment models where extremism
increases due to interactions between radicals and non-radicals
\cite{CAS03,DEU15,SHO17}. By invoking the Bendixson-Dulac theorem
\cite{BUR05,DUM06} with $1 / \left[ n_2 \left( 1 - n_0 - n_2 \right)
\right]$ as the Dulac function, it can be shown that limit cycles do
not arise. All trajectories converge to fixed points $(n_0^*, n_2^*)$
as shown in Figs.\,\ref{FIG:NO_AGE} and \ref{FIG:NO_AGE1}. Note that
all fixed points must satisfy the constraint $0 \leq n_i^* \leq 1$ for
all $i$.  Sociologically, within some radical groups pacification may
be rare due to extreme die-hard fanaticism, while in other groups
gradually retracting from radicalization may be possible.  The
reversibility of the final, radical stage (i.e., whether $c_{\rm P}=0$
or $c_{\rm P} > 0$) plays a key role in determining qualitative
behaviors as observed in opinion dynamics models \cite{VER14,WAA15}.
Hence in analyzing Eqs.\,\ref{EQ:NONACTIVIST1} and \ref{EQ:RADICAL1}
we consider two distinct cases: reversible radicalization, whereby
$c_{\rm P} > 0$, and irreversible radicalization, whereby $c_{\rm P} =
0$.

If $c_{\rm P} > 0$ and radicalization is reversible, a maximum of
three fixed points may arise, as shown in
Figs.\,\ref{FIG:NO_AGE}(a)--(c). The fixed point at $( n_0^*=1,
n_2^*=0 )$ (and $n_1^* = 0$), where all individuals are non-radical.
This fixed point corresponds to ``utopia", a society without any
activists or radicals.  The second fixed point is the ``dormant" state
$(n_0^* = c_{\rm D} / c_{\rm A}, n_2^* = 0 )$ (and $n_1^* = 1 - c_{\rm
  D}/c_{\rm A})$, characterized by the presence of activists who do
not turn radical.  Here the non-radical population $n_0^*$ is given by
balancing the $c_{\rm D}$ de-activation-induced influx and the $c_{\rm
  A}$ activation-induced efflux, yielding $n_0^* = c_{\rm D}/c_{\rm
  A}$.  Finally, the third ``turmoil" fixed point is at $(n_0^* =
c_{\rm D}/c_{\rm A}, n_2^* = 1 - c_{\rm D}/c_{\rm A} - c_{\rm
  P}/c_{\rm R})$ (and $n_1^* = c_{\rm P}/c_{\rm R}$), where a non-zero
fraction of the population is permanently radicalized.  While the
fraction of the non-radical population is again given by $n_0^* =
c_{\rm D}/c_{\rm A}$, the activist population is similarly determined
by balancing the $c_{\rm P}$ pacification-induced influx and the
radicalization-induced efflux $c_{\rm R}$, yielding $n_1^* = c_{\rm
  P}/c_{\rm R}$.  The existence and stability of these fixed points
depend on the values of $c_{\rm D}/c_{\rm A}$ and $c_{\rm P}/c_{\rm
  R}$ as well as whether they satisfy the $0 \leq n^*_i \leq 1$
constraints.

When de-activation dominates activation and $c_{\rm D}/ c_{\rm A} > 1$
utopia is the only feasible steady state.  The phase portrait of the
system is shown in Fig.\,\ref{FIG:NO_AGE}(a): all trajectories
converge to $(1,0)$. Upon increasing activation so that $ 1 - c_{\rm
  P}/c_{\rm R} < c_{\rm D} /c_{\rm A} \leq 1$, the dormant state
enters the physical domain and becomes the only stable solution. The
phase portrait in Fig.\,\ref{FIG:NO_AGE}(b) shows that all
trajectories converge to $(c_{\rm D} / c_{\rm A}, 0)$, while $(1, 0)$
becomes a saddle point. Finally, when activation dominates
de-activation and $c_{\rm D}/c_{\rm A} \leq 1 - c_{\rm P}/c_{\rm R}$,
turmoil replaces the dormant state as the only stable solution.
Fig.\,\ref{FIG:NO_AGE}(c) shows all trajectories converging to $(
c_{\rm D}/c_{\rm A}, 1 - c_{\rm D}/c_{\rm A} - c_{\rm P}/c_{\rm R})$
while the other two fixed points become saddle nodes.

%
%

Under irreversible radicalization, when $c_{\rm P} = 0$, in addition
to the above fixed points, a family of solutions $n_0^* + n_2^* = 1$
(and $n_1^* = 0$) will arise. This follows from the irreversible
nature of radicalization: in the absence of a pacification mechanism
the $n_2$ radical group becomes a sink, eventually depleting the $n_1$
activist pool.  As a consequence, population exchanges between $n_0$
and $n_2$ are interrupted, stranding phase trajectories along the
$n_0^* + n_2^* = 1$ line.  Note that when $c_{\rm P} = 0$, the
stability interval for the dormant state, $1 - c_{\rm P}/c_{\rm R} <
c_{\rm D} /c_{\rm A} \leq 1$ vanishes, so that the dormant state is
always unstable.  Under irreversible radicalization, increases in
$c_{\rm A}$ shift the system from utopia directly to turmoil.

In Fig.\,\ref{FIG:NO_AGE1}(a) we plot the phase portrait for $c_{\rm
  D}/ c_{\rm A} >1$; the fixed point $(1,0)$ and the solution family
$n_0^* + n_2^* = 1$ are both marginally stable.  In
Figs.\,\ref{FIG:NO_AGE1}(b) and (c) we set $c_{\rm D}/ c_{\rm A} \leq
1$ and consider different values of $c_{\rm R}$. In both figures, the
fixed point $\left( c_{\rm D}/c_{\rm A}, 1 - c_{\rm D}/c_{\rm A}
\right)$ is marginally stable while $(1, 0)$ is unstable. The segments
of the solution family $n_0^* + n_2^* = 1$ between $(1,0)$ (open
circle) and $\left((c_{\rm D} + c_{\rm R})/(c_{\rm A} + c_{\rm R}), 1
- (c_{\rm D} + c_{\rm R})/(c_{\rm A} + c_{\rm R}) \right)$ (open star)
become unstable.  Due to the irreversibility of radicalization the
dynamics are highly sensitive to initial conditions; where a
trajectory will end along the stable portion of the $n_0^* + n_2^* =
1$ line depends on where the trajectory starts.  An important role is
also played by the radicalization rate $c_{\rm R}$. As the latter
increases, the open star $\left( (c_{\rm D} + c_{\rm R})/(c_{\rm A} +
c_{\rm R}), 1 - (c_{\rm D} + c_{\rm R})/(c_{\rm A} + c_{\rm R})
\right)$ will shift towards $(1, 0)$, expanding the stable portion of
the solution family.  Larger $c_{\rm R}$ values, as shown in
Fig.~\ref{FIG:NO_AGE1}(c), enhance the stability of the solution
family, as trajectories are steered more vigorously towards $n^*_0 +
n^*_2 = 1$, before turmoil can be reached.

\begin{figure} [h]
  \begin{center}
      \includegraphics[width=2in]{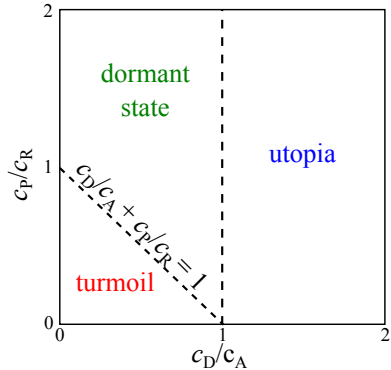}
\end{center}
  \caption{Phase diagram of the age-independent model under reversible
    radicalization. For $c_{\rm D} \ge c_{\rm A}$, the system settles
    at utopia $(n_0^* =1, n_1^* = 0, n_2^*= 0)$. For $c_{\rm D} <
    c_{\rm A}$ and $c_{\rm D}/c_{\rm A} + c_{\rm P}/c_{\rm R} \ge 1$,
    the dormant state is stable at $(n_0^* = c_{\rm D}/c_{\rm A},
    n_1^*= 1- c_{\rm D}/c_{\rm A}, n_2^*=0)$. For $c_{\rm D}/c_{\rm A}
    + c_{\rm P}/c_{\rm R} < 1$, the system is in turmoil at $(n_0^* =
    c_{\rm D}/c_{\rm A}, n_1^*=c_{\rm P}/c_{\rm R}, n_2^*= 1- c_{\rm
      D}/c_{\rm A} - c_{\rm P}/c_{\rm R})$.
         \label{FIG:PHASE_DIAGRAM} }
\end{figure}

The above analysis reveals that three steady states are possible
within the age-independent formulation of model: utopia, a dormant
state, and turmoil, as summarized in Fig.\,\ref{FIG:PHASE_DIAGRAM}.
Under reversible radicalization, the system settles into one of the
three depending on the values of $c_{\rm D}/c_{\rm A}$ and $c_{\rm P}/
c_{\rm R}$, while for irreversible radicalization the final steady
state also depends on initial conditions.

\subsection{Full age-dependent model}

\noindent
We now consider the full age-dependent model as defined in
Eqs.\,\ref{EQ:RHO0}--\ref{kernels}, starting with the scenario of
reversible radicalization, $C_{\rm P} > 0$.  As shown in
Fig.\,\ref{FIG:NO_AGE} the corresponding age-dependent model
progresses through the utopia, dormant, and turmoil steady states as
$c_{\rm D}$ decreases.  In order to compare results, we begin by
examining the steady states arising from the age-dependent model by
varying $C_{\rm D}$ while fixing other parameters. We assume the
radicalization window to occur between ages $a_0=5$ and $a_1 = 55$; to
include age dependence we must also specify $\alpha_j$ and $\sigma_j$.
For simplicity we assume the same age dependence for activation and
radicalization, setting $\alpha_{\rm A} = \alpha_{\rm R} = 20$ years,
and $\sigma_{\rm A} = \sigma_{\rm R} = 10$ years.  We assume initial
conditions $\rho_0(a, 0) = 0.989$, $\rho_1(a, 0) = 0.01$, and
$\rho_2(a, 0) = 0.001$, representing an overwhelmingly non-radical
society. Finally, we choose $C_{\rm A} = C_{\rm R} = 2$, $C_{\rm P} =
0.5$.  The initial conditions yield a total population $N = \bar{a}$,
leading to the corresponding parameter values $c_{\rm A} = c_{\rm R} =
1$, $c_{\rm P} = 0.5$ for the age-independent model, which are exactly
the same values used in Fig.\,\ref{FIG:NO_AGE}.  Since $c_{\rm D} =
C_{\rm D}$ we finally select $C_{\rm D}= 2$, $0.6$, and $0.2$, so that
full comparisons can be made with Fig.\,\ref{FIG:NO_AGE} and the
implications of including age dependence in the model can be
highlighted.  Within the age-dependent scenario analytical estimates
of steady states are not available and numerical simulations must be
run for every set of chosen parameters.  Henceforth, all times and age
units will be assumed to be years.  All main parameters and terms are
summarized in Table \ref{TABLE1}.

\begin{figure*}[t]
  \begin{center}
      \includegraphics[width=6.9in]{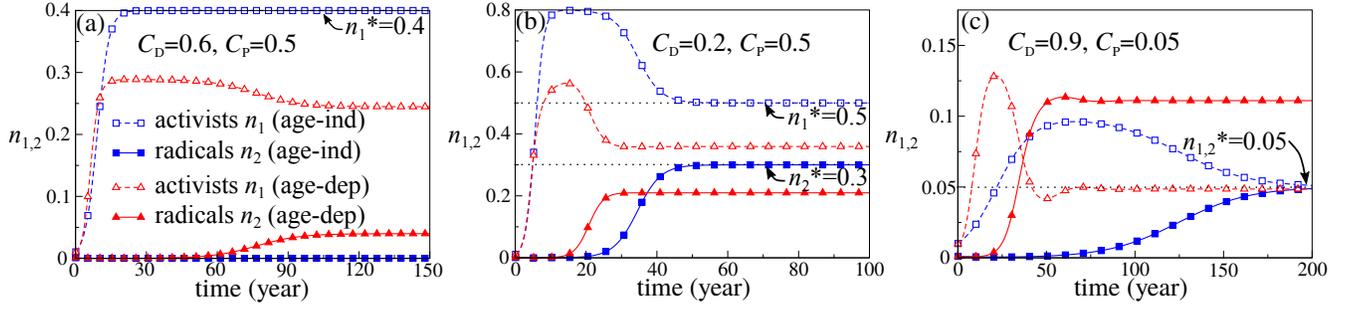}
\end{center}
  \caption{ Time dependence of the the activist $n_1$ and radical
    $n_2$ population fractions from the age-independent and
    age-dependent formulations of our model. We fix $C_{\rm A} =
    C_{\rm R} = 2$ in all panels and set $C_{\rm D} = 0.6$ and $C_{\rm
      P} = 0.5$ in panel (a); $C_{\rm D} = 0.2$ and $C_{\rm P} = 0.5$
    in panel (b) and $C_{\rm D} = 0.8$ and $C_{\rm P} = 0.1$ in panel
    (c).  Additional parameters for the age-dependent model are set to
    $\alpha_{\rm A} = \alpha_{\rm R} = 20$, $\sigma_{\rm A} =
    \sigma_{\rm R} = 10$.  Initial conditions are $(\rho_0(a, 0),
    \rho_1 (a, 0), \rho_2 (a, 0)) = (0.989, 0.01, 0.001)$.  In panel
    (a), age dependence allows a small fraction of radicals to rise,
    while the age-independent model predicts a dormant state at
    $(n_0^* = 0.6, n_1^* = 0.4, n_2^* = 0)$. In panels (b) and (c),
    both models evolve towards turmoil.  In panel (b) the
    age-independent model predicts aggressive radicalization with
    turmoil at $(n_0^* = 0.2, n_1^* = 0.5, n_2^* = 0.3)$; including
    age-structure leads to a lower radical fraction. Radicalization is
    less aggressive in panel (c). Here, the age-independent turmoil
    steady state is at $(n_0^* = 0.9, n_1^* = 0.05, n_2^* = 0.05)$ and
    age dependence increases the radical population. Note that in all
    panels, age dependence accelerates activist and radical growth,
    regardless of the final steady states.
       \label{FIG:CP_POSITIVE} }
\end{figure*}

In Fig.\,\ref{FIG:CP_POSITIVE}, we numerically evaluate
Eqs.\,\ref{EQ:RHO0}--\ref{kernels} for $t=100$ and plot the $n_1(t)$
activist and $n_2(t)$ radical populations as defined by
Eqs.\,\ref{nidef}.  For comparison, we also show the corresponding
age-independent $n_1(t), n_2(t)$ populations obtained from
Eqs.\,\ref{EQ:NONACTIVIST}--\ref{EQ:RADICAL} where initial conditions
were specified as $n_0(0) =0.989$, $n_1 (0) = 0.01$, and $n_2 (0) =
0.001$ for consistency.  For $C_{\rm D} = 2$ the age-independent model
predicts utopia, a result that is confirmed in the age-dependent
case. We do not plot the dynamics corresponding to this case, since
both $n_1 (t)$ and $n_2 (t)$ quickly vanish in both formulations.  For
$C_{\rm D} = 0.6$, Fig.\,\ref{FIG:CP_POSITIVE}(a) shows that a small
but nonzero radical fraction $n_2(t \to \infty)$ emerges within the
age-dependent model, in contrast to the age-independent prediction of
a dormant state whereby $n_2^*=0$.  This discrepancy is due to
age-enhancement of the activation $A(a; \rho_1)$ and radicalization
$R(a; \rho_2)$ rates near the early ages $a \simeq \alpha_j$ for $j
={\rm A,R}$ respectively. Recall, that by construction, and as
discussed when justifying the shape of the sensitivity kernels in
Eq.\,\ref{kernels}, enhancing the transition rates at a given age,
necessarily decreases them at other age intervals.  Here, increasing
the transition rates around $\alpha_{j}$ is sufficient to drive the
system towards turmoil, despite the same transition rates being
lowered in other age windows.

\begin{table}[t]
\begin{tabular}{|c|c | c |}
\hline
Symbol & Description & Values \\
\hline
 $\, \, A (a; \rho_1) \, \,$ & activation rate coefficient & Eq.~(\ref{EQ:ACTIVATION_RATE})  \\ 
\hline
 $\, \, R (a; \rho_2) \, \,$ & radicalization rate coefficient & Eq.~(\ref{EQ:RADICALIZATION_RATE})  \\ 
\hline
 $G$ & government conversion rate coefficient & Eq.~(\ref{EQ:GOV})  \\ 
\hline
 $C_{\rm A}$ & activation intensity & 1 -- 12 \\
\hline
 $C_{\rm R}$ & radicalization intensity & 0 -- 20 \\
\hline
 $C_{\rm D}$ & de-activation intensity & 0.2 -- 5 \\
\hline
 $C_{\rm P}$ & pacification intensity & 0 -- 0.5 \\
\hline
 $C_{\rm G}$ & government intervention intensity & 0 - 5 \\
\hline
 $\alpha_{\rm A}$ & activation targeted age & 20 ys.\\
\hline
 $\alpha_{\rm R}$ & radicalization targeted age & 20 ys. \\
\hline
 $\alpha_{\rm G}$ & government targeted age & 20 ys.\\
\hline
 $a_0$ & lower bound of interaction age window & 5 ys. \\
\hline
 $a_1$ & upper bound of interaction age window & 55 ys.\\
\hline
 $\sigma_{\rm A}$ & activation age breadth & 10 ys. \\
\hline
 $\sigma_{\rm R}$ & radicalization age breadth & 10 ys. \\
\hline
 $\sigma_{\rm G}$ & government intervention age breadth & 10 ys. \\
\hline
 $\mu$ & fraction of intervention on activists & 0 -- 1 \\
\hline
\end{tabular}
\begin{center}
\caption{List of parameters and functional forms used in the model
  presented in Eqs.\,\ref{EQ:RHO0}--\ref{kernels}. Government
  intervention quantities are discussed in Section \ref{govt}.
\label{TABLE1}}
\end{center}
\end{table}

Finally, for $C_{\rm D} = 0.2$ both the age-dependent and
age-independent models reach turmoil as shown in
Fig.\,\ref{FIG:CP_POSITIVE}(b). Note that the steady-state activist
and radical fractions arising from the age-dependent model, $n_1(t \to
\infty) = 0.36$ and $n_2(t \to \infty) = 0.21$ are lower than the
predicted $n_1^* = 0.5$ and $n_2^* = 0.3$ values in the
age-independent case; this is due to the age-structure of the kernel
in Eq.\,\ref{kernels}.  In the age-independent case, where the
transition rates $c_{\rm A}$ and $c_{\rm R}$ are not age-structured,
the chosen parameters lead to turmoil where only a relatively small
population is non-radical: $n_0^* = 1- n_1^* - n_2^*= 0.2$. The
age-dependent formulation allows for age variability in the transition
rates $A(a; \rho_1)$ and $R(a; \rho_2)$, which increase at the peak
ages $\alpha_{\rm A}$ and $\alpha_{\rm R}$, but decrease for other
ages. The local increases in $A(a \simeq \alpha_{\rm A}; \rho_1)$ and
$R(a \simeq \alpha_{\rm R}; \rho_2)$ do not appreciably raise the
number of activists or radicals compared to those in the turmoil state
of the age-independent formulation, since they are already
sustained. Decreasing the transitions in other age-windows, however,
stymies the radicalization of other individuals far from the peak age,
resulting in an overall net decrease of the radical and activist
populations so that $n_1(t \to \infty) < n_1^*$ and $n_2(t \to \infty)
< n_2^*$.

In Fig.\,\ref{FIG:CP_POSITIVE}(c) we select $C_{\rm A} = C_{\rm R} =
2$, $C_{\rm D} = 0.9$, and $C_{\rm P} = 0.05$ to yield the
age-independent, turmoil steady state $(n_0^*=0.9, n_1^* = 0.05, n_2^*
= 0.05)$, with a large non-radical population. In the age-dependent
model, enhanced rates at the peak ages $\alpha_{\rm A}$ and
$\alpha_{\rm R}$ allow for the transition of a much larger non-radical
population into the activist and radical pools, compared to the case
discussed in Fig.\,\ref{FIG:CP_POSITIVE}(c) where the non-radical
population was small. Overall, a larger value for $n_2 (t \to \infty)=
0.11$ is observed in the age-dependent formulation than in the
age-independent one.  Note that in both
Figs.\,\ref{FIG:CP_POSITIVE}(b) and (c), activists and radicals emerge
earlier in the age-dependent than in the age-independent model,
regardless of the final steady-state. This trait persists for several
parameter choices, suggesting that age-structured rates accelerate
radicalization, due to their enhancement at young ages.

\begin{figure*}[t]
  \begin{center}
      \includegraphics[width=6.9in]{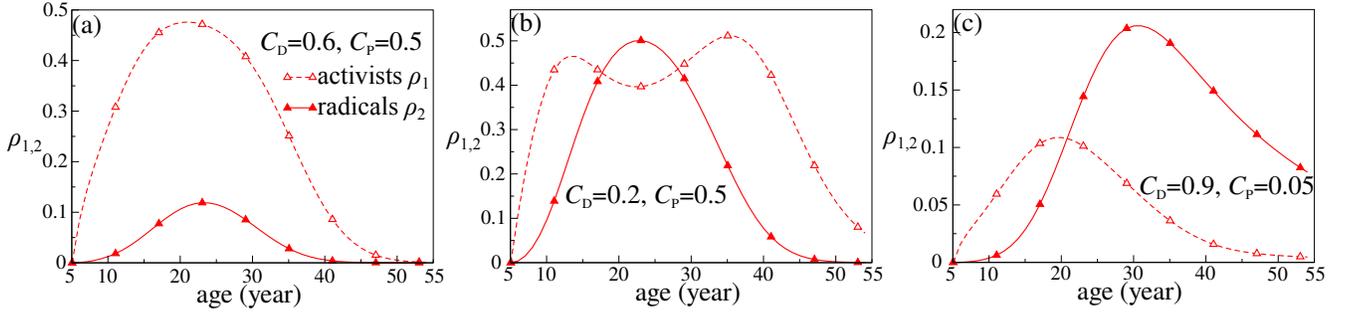}
\end{center}
  \caption{Age distribution of activists and radicals for the
    parameters used in Figs.\,\ref{FIG:CP_POSITIVE}(a)--(c).  Panel
    (a) shows a small fraction of radicals, deviating from the dormant
    steady state predicted by the age-independent model.  Both
    activist and radical populations peak at ages slightly larger than
    $\alpha_{\rm R}$.  Panel (b), corresponding to strong turmoil,
    shows the radical population peaking at ages slightly larger than
    $\alpha_{\rm R}$, accompanied by a bi-modal age-distribution of
    activists. Panel (c) corresponds to mild turmoil and shows a
    uni-modal age distribution for both radicals and activists, with
    the radical peak occurring at ages larger than $\alpha_{\rm A}$
    and the activist maximum at ages lower than $\alpha_{\rm A}$. The
    mechanisms that lead to uni-modal or bimodal distributions as well
    as to the shifts in the peaks compared to $\alpha_{\rm A}$ are
    detailed in the text.
         \label{FIG:CP_POSITIVE2} }
\end{figure*}

In Fig.\,\ref{FIG:CP_POSITIVE2} we plot the age-structured
distributions for activists and radicals $\rho_1(a, t \to \infty)$ and
$\rho_2 (a,t \to \infty)$ as approximated at $t=100$ after steady
state has been reached.  In Fig.\,\ref{FIG:CP_POSITIVE2}(a) we use the
same parameters as in Fig.\,\ref{FIG:CP_POSITIVE}(a) where the
age-independent model predicts a dormant state.  Here, the
age-structured steady state population is mostly in the activist
state, peaking at age $a \simeq 21$, although a small fraction of
radicals $\rho_2 (a, t \to \infty)$ emerges and peaks at age $a \simeq
23$, due to the locally increased transition rates, as described
above.  Note that the maximum in $\rho_{1}(a, t \to \infty)$ occurs at
age $a^* > \alpha_{\rm A}$.  This can be explained by noting that the
greatest transition towards the activist state occurs at age
$\alpha_{\rm A}$, when the sensitivity kernel ${\cal
  K}(a,a',\alpha_{\rm A}, \sigma_{\alpha})$ reaches its maximum.  The
individuals who joined the activist group at age $\alpha_{\rm A}$ will
age and remain, at least for some time, in this state before
transitioning to the radical or non-radical states for ages $a >
\alpha_{\rm A}$.  Additional individuals will join the activist pool
at older ages, yielding a cumulative activist population that is
greatest for $a^* > \alpha_{\rm A}$, as $\rho_1(a^*, t \to \infty)$
contains individuals that transitioned earlier and aged while
remaining in the activist state, but also those who transitioned
later.  At older ages, the kernel in Eq.\,\ref{kernels} decreases:
individuals leave the activist pool in larger numbers than they will
join it, so that $\rho_1(a, t \to \infty)$ decreases from its maximum.
A similar argument can be applied to $\rho_2(a, t \to \infty)$ which
also peaks at age $a > \alpha_{\rm R}$.

%
%

A more interesting case is shown in Fig.\,\ref{FIG:CP_POSITIVE2}(b)
where the chosen parameters are the same as in
Fig.\,\ref{FIG:CP_POSITIVE}(b) and where the de-activation rate $C_{\rm
  D}$ is lower than what used in Fig.\,\ref{FIG:CP_POSITIVE2}(a).  In
this case, the corresponding age-independent model reaches a strong
turmoil steady state.  We observe that the radical population in
Fig.\,\ref{FIG:CP_POSITIVE2}(b) peaks near $a \simeq 23$, while
activists are bimodally distributed, peaking at both younger $a \simeq
13$, and older ages $a \simeq 35$.  Of the two $\rho_1(a; t \to
\infty)$ maxima, the first occurs at age $a < \alpha_{\rm A}$.  Here,
due to the low value of $C_{\rm D}$, a large activist population
rapidly emerges at young ages; these individuals gradually transition
towards the radical stage, and further increase $R(a;
\rho_2)$. Eventually, the radicalized population becomes so numerous
that $R(a; \rho_2)$ overtakes the activation rate $A(a; \rho_1)$,
causing activists to vigorously transition to the radical pool,
effectively leading to the net loss of activists around age $a \simeq
\alpha_{\rm A}$ as seen in Fig.\,\ref{FIG:CP_POSITIVE2}(b).  The first
peak in $\rho_1(a, t \to \infty)$ thus emerges at younger ages $a <
\alpha_{\rm A}$, when the number of radicals is relatively low and so
is $R(a; \rho_2)$. At intermediate ages $a > \alpha_{\rm A}$, $R(a;
\rho_2)$ begins to decline along with the radical population
$\rho_2(a, t \to \infty)$, leading to an increase in the activist
population $\rho_1(a, t \to \infty)$. These individuals shift to the
non-activist pool at a slow rate as $C_{\rm D}$ is relatively low. As
older ages are approached activation further decreases, $\rho_1(a, t
\to \infty)$ begins to decline and individuals increasingly return to
the non-radical state. These two competing age-dependent mechanisms
give rise to the second peak in $\rho_1(a; t \to \infty)$ as seen in
Fig.\,\ref{FIG:CP_POSITIVE2}(b).

Finally, in Fig.\,\ref{FIG:CP_POSITIVE2}(c) we show the age dependence
of the steady state radical and activist populations using the same
parameters as in Fig.\,\ref{FIG:CP_POSITIVE}(c), where the
age-independent model predicts mild turmoil. We find that activists
peak at age $a^* \simeq 19$, while radicals reach their maximum at age
$a^* \simeq 28 $.  The mechanisms driving the observed age-dependent
patterns are similar to what described for
Fig.\,\ref{FIG:CP_POSITIVE2}(b), however here we do not observe a
secondary peak for $\rho_2(a, t \to \infty)$.  In this case the
de-activation rate $C_{\rm D} = 0.9 $ is larger than what used in
Fig.\,\ref{FIG:CP_POSITIVE2}(b), where $C_{\rm D} = 0.2$.  As a
result, as the radicalization rate $R(a; \rho_2)$ decreases at older
ages, the large de-activation rate does not allow for an intermediate
activist population to persist, rather most of the former radicals
turn directly to non-radical behavior.  The activist population
declines continuously in age and does not exhibit a secondary maximum.
Upon comparing Figs.\,\ref{FIG:CP_POSITIVE2}(b) and (c) we also see
that while in the state of strong turmoil, adherents may remain active
even at older ages when the number of radical extremists vanishes. On
the other hand, panel (c) shows that in cases of mild turmoil, the
number of radicals and activists is concentrated at early ages,
suggesting that fervor wanes significantly with age.

We now consider the case of irreversible radicalization $C_{\rm P}=0$
in Eqs.\,\ref{EQ:RHO0}--\ref{kernels}.  In the previous subsection we
showed that in the age-independent case, the dynamics is strongly
dependent on the radicalization rate; hence we begin by illustrating
the particular case of fixed $C_{\rm A} = 12$, $C_{\rm D} = 5$, and
varying $C_{\rm R}$.  We assume the same age parameters as the
reversible case above, where $\alpha_{\rm A} = \alpha_{\rm R} = 20$,
and $\sigma_{\rm A} = \sigma_{\rm R} = 10$.  Initial conditions are
also chosen as above with $\rho_0(a, 0) = 0.989$, $\rho_1(a, 0) =
0.01$, and $\rho_2(a, 0) = 0.001$.  Within the age-independent model,
the above parameters correspond to $c_{\rm A} = 6$, $c_{\rm D} = 2.5$
and $c_{\rm R} = C_{\rm R}/2$ and yield a stable turmoil state at
$(n_0^*=5/6, n_1^*=0 , n_2^* = 1/6)$ and an unstable dormant state at
$(n_0^*=5/6, n_1^*=1/6, n_2^*=0)$; neither of these configurations
depend on $c_{\rm R}$.

%
%

\begin{figure*}[t]
  \begin{center}
      \includegraphics[width=6.9in]{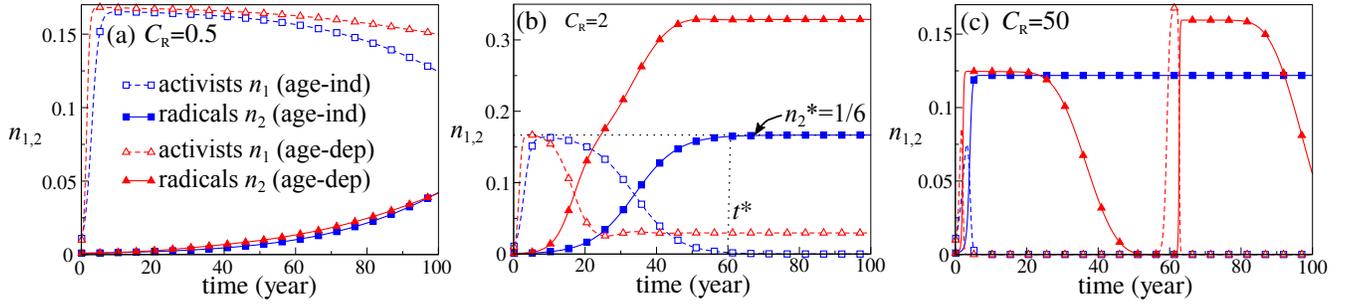}
\end{center}
  \caption{Time dependence of the activist $n_1$ and radical $n_2$
    populations from the age-independent and age-dependent
    formulations of our model. We fix $C_{\rm A} = 12$ and $C_{\rm D}
    = 5$ and set $C_{\rm R} = 0.5$ in panel (a), $C_{\rm R} = 2$ in
    panel (b), and $C_{\rm R} = 50$ in panel (c).  Additional
    parameters for the age-dependent model are $\alpha_{\rm A} =
    \alpha_{\rm R} = 20$, $\sigma_{\rm A} = \sigma_{\rm R} = 10$.
    Initial conditions are $(\rho_0(a, 0), \rho_1 (a, 0), \rho_2 (a,
    0)) = (0.989, 0.01, 0.001)$.  In all panels the activist
    population rises first, followed by radical growth. In the
    age-independent case, $n_1(t), n_2(t)$ saturate either at turmoil
    as shown in panels (a) and (b) or along the solution family as
    shown in panel (c).  Age-dependence slightly accelerates the
    growth of activists and radicals at low $C_{\rm R}$ in panel (a);
    greatly enhances the growth of radicals by maintaining a finite
    activist population at intermediate $C_{\rm R}$ in panel (b); and
    leads to activist and radical population oscillations at high
    $C_{\rm R}$ in panel (c).}
\label{FIG:AGE_TIME}
    \end{figure*}

In Figs.\,\ref{FIG:AGE_TIME}(b) we increase $C_{\rm R} =2$ ($c_{\rm R}
= 1$), while keeping all other parameters unchanged. In the
age-independent formulation the system converges to turmoil at $(n_0^*
= 5/6, n_1^* = 0, n_2^* = 1/6)$ after roughly 60 years. The dynamics
are quite different in the age-dependent version of the model. Here,
the activist fraction $n_1(t)$ rises and declines more abruptly than
in the age-independent case; similarly the radical fraction $n_2(t)$
increases at earlier times and more vigorously, greatly surpassing the
$n^*_2 = 1/6$ steady state value of the age independent model at long
times.  Of particular importance is the fact that age-dependent
transitions allow the $n_1(t)$ activist population to exist
continuously and not vanish as it does in the age-independent case.
This non-depleted pool of $n_1(t)$ radicals provides an uninterrupted
source of recruits for $n_2(t)$, which can grow to large values; it
also provides a pathway for more $n_0(t)$ non-radicals to be initiated
into the radicalization process.  Indeed, $n_0 (t \to \infty)= 0.64$
at steady state in the age-dependent model as opposed to $n^*_0 = 5/6
= 0.83$ in the age-independent case.  These results show that age
structure allows for a broad influence of the activist population:
age-differentiation allows activists to persist over different age
windows and not be depleted all at once by joining the radical
pool. Thus, while still maturing towards full fledged radicalization,
activists keep recruiting members into their ranks through
peer-pressure, continuously funneling individuals to the radical
state.

If we further increase $C_{\rm R}= 50$ ($c_{\rm R} = 25$), as shown in
Figs.~\ref{FIG:AGE_TIME}(c) we observe that, due to the very large
radicalization rate, the activist fraction $n_1(t)$ is quickly
depleted even within the age-dependent model. The loss of the
intermediate activist population represents a bottleneck hindering
radical population growth, as new members are no longer being
recruited.  As a result, even with a higher radicalization rate, the
$n_2 (t \to \infty)$ steady states in this case are lower than in
Fig.\,\ref{FIG:AGE_TIME}(b) both in the age-dependent and
age-independent models.  In particular, in the age-independent
formulation, the system will irreversibly settle along the $n_0^* +
n_2^* = 1$ solution family, and not at the turmoil state.  On the
other hand, as can be seen in the dynamics of the age-dependent
$n_2(t)$ in Fig.\,\ref{FIG:AGE_TIME}(c) the continuous presence of
radicals becomes unsustainable without activists supplying new
recruits.  Here, radicals eventually fade away due to old age and are
replaced by non-radicals, under the population conservation conditions
assumed in our model.  This depletion slows the further radicalization
of activists, who interact less with full fledged radicals, allowing
$n_1(t)$ to recover, as can be seen in the resurgent spike of
activists in Fig.\,\ref{FIG:AGE_TIME}(c). The rise in $n_1(t)$
triggers the regrowth of $n_2(t)$, which once again depletes the
activist pool, leading to oscillations in $n_1(t)$ and $n_2(t)$ over
time. During the second radicalization wave (and other hypothetical
ones beyond the $100$ years considered in this panel) the $n_1(t)$,
$n_2(t)$ populations reach higher values than during the first,
transient one.  Interestingly, the emergence of activists is seen to
precede that of radicals by roughly five years, on the same time-scale
as reported for a demographic study of individuals progressing from
initial radicalization to full extremism \cite{KLA16}.

%
%

\begin{figure}[h]
  \begin{center}
      \includegraphics[width=2.32in]{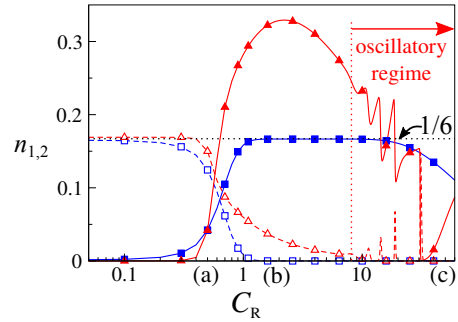}
\end{center}
  \caption{ Long-time $n_1(t=100)$ and $n_2(t=100)$ populations as a
    function of $C_{\rm R}$ ($c_{\rm R} = C_{\rm R}/2$) for the
    age-independent and age-dependent models.  The $C_{\rm R}$ values
    used in Fig.\,\ref{FIG:AGE_TIME} are explicitly marked. At low
    $C_{\rm R}$, the radical population $n_2$ is negligible, and the
    activist population $n_1$ saturates.  At intermediate $C_{\rm R}$,
    the $n_2$ radicals rise by converting more $n_1$ activists.  Note
    that $n_1(t=100) \to 0$ in the age-independent case, while in the
    age-dependent model $n_1(t=100)$ remains finite and leads to an
    enhanced radical population. At high $C_{\rm R}$, the
    age-structured model yields oscillatory dynamics as shown by the
    spiky curves. In the age-independent model, $n_2(t = 100)$
    declines as the system evolves towards the solution family $n_0^*
    + n_2^* =1$ and away from turmoil at $n_2^* = 1/6$.}
    \label{FIG:AGE_TIME2}
\end{figure}

In Fig.\,\ref{FIG:AGE_TIME2}, we plot long term population fractions
$n_i(t=100)$ for $i=1,2$ as a function of $C_{\rm R}$ ($c_{\rm R} =
C_{\rm R}/2$) for the two formulations of our model. For low $C_{\rm
  R} \lesssim 1$, radical fractions $n_2(t=100)$ are negligible while
activist populations $n_1(t=100)$ remain close to metastable
levels. These are slightly higher in the age-dependent case than in
the age-independent one, as discussed above.  Increasing
radicalization rates to moderate $1 \lesssim C_{\rm R} \lesssim 10$
values drives the age-independent model to the $c_{\rm R}$-independent
turmoil steady state at $(n_0^* = 5/6, n_1^*=0, n_2^* = 1/6)$, where
$n_1^*$ vanishes. In contrast, within the age-dependent model
$n_1(t=100)$ decreases with increasing $C_{\rm R}$ but stays non-zero,
leading to a significantly elevated $n_2(t=100)$ as discussed
above. We find that there exists an optimal $C^{\rm opt}_{\rm R}$
value that maximizes $n_2(t=100)$ in the age-dependent case.  Values
of $C_{\rm R}$ that are too low hinder recruitment from the activist
pool; values of $C_{\rm R} $ that are too large overtax, it leading to
an optimal $C^{\rm opt}_{\rm R}$ value.  As $C_{\rm R} \gtrsim 10$ is
further increased, $n_1(t=100)$ continues to decline in the
age-dependent model. Eventually the oscillatory regime is reached, as
shown in Fig.\,\ref{FIG:AGE_TIME2}.  The age-independent model does
not exhibit oscillations for large $C_{\rm R}$, however $n_2^*$
declines, indicating that turmoil is no longer reached, and that the
system settles along the solution family $n_0^* + n_2^* = 1$ as
described at the end of subsection \ref{AI}.

%
%

%
\begin{figure*}[t]
  \begin{center}
      \includegraphics[width=6.9in]{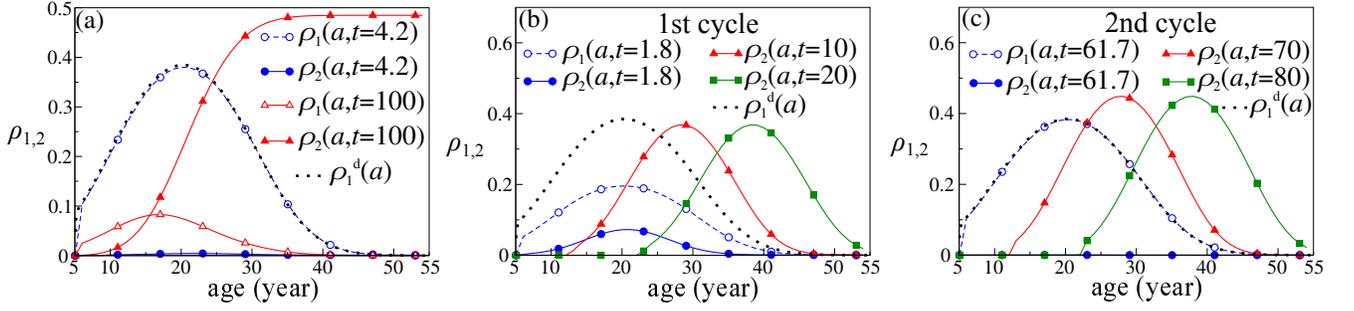}
\end{center}
  \caption{Age distribution of activists and radicals for the
    parameters used in Figs.\,\ref{FIG:AGE_TIME}(b) and (c).  In panel
    (a) activists emerge first and at $t = 4.2$ their age distribution
    is close to the dormant distribution $\rho_1^{\rm d}(a)$. At long
    times ($t = 100$ years) radicals surpass the activist population.
    Their age distribution increases at young ages and plateaus at
    older ones. A small fraction of activists remains, peaking at ages
    younger than $\alpha_{\rm A}$.  Panel (b) depicts the age
    distributions during the transient phase of the oscillatory
    regime. At $t=1.8$ activists reach a maximum, and a small radical
    fraction emerges. At $t=10$ and $20$ the activist population (not
    plotted) vanishes, while that of radicals increases. The age
    distribution of radicals maintains its shape as it advects towards
    older ages.  Panel (c) shows the age distributions of the
    oscillatory regime after the transient period. Behaviors are
    qualitatively similar to panel (b). However, when activists reach
    a maximum at $t=61.7$ years, the radical population remains
    negligible, allowing activists to approach the dormant state and
    increasing the radical population at $t=70$ and $80$.
         \label{FIG:CP0_AGE_DIST} }
\end{figure*}
%

In Fig.\,\ref{FIG:CP0_AGE_DIST} we show the age distribution of
activists and radicals at long times ($t = 100$) under irreversible
radicalization $C_{\rm P} = 0$ using the same parameters as in
Figs.\,\ref{FIG:AGE_TIME}(b) and (c).  We do not display the age
distribution profiles corresponding to Fig.\,\ref{FIG:AGE_TIME}(a)
since the system is not equilibrated at $t=100$ and the age
distribution curves would be qualitatively similar to the short time
ones corresponding to Fig.\,\ref{FIG:AGE_TIME}(b).

 In Fig.\,\ref{FIG:CP0_AGE_DIST}(a) we use the same parameters as in
 Fig.\,\ref{FIG:AGE_TIME}(b).  For the activist age distribution
 $\rho_1(a, t \to \infty)$ a peak arises at age $a^* < \alpha_{\rm A}$
 and the distribution is similar to what observed in
 Fig.\,\ref{FIG:CP_POSITIVE2}(c) under reversible
 radicalization. Although the parameters are different, the mechanism
 leading to the early maximum in Fig.\,\ref{FIG:CP0_AGE_DIST}(a) is
 the same as what outlined when discussing
 Fig.\,\ref{FIG:CP_POSITIVE2}(c): namely activists quickly joining the
 rank of radicals as they mature, yielding an early age peak.  In
 contrast, due to the irreversible nature of radicalization, the age
 distribution of radicals $\rho_2(a, t \to \infty)$ increases
 continuously until a plateau is reached since, once radicalized,
 individuals can only age within the extremist pool but not
 deradicalize.

In Fig.\,\ref{FIG:CP0_AGE_DIST}(a) we also plot the age distributions
at $t = 4.2$ when the fraction of activists $n_2(t=4.2)$ is at the
maximum value shown in Fig.\,\ref{FIG:AGE_TIME}(b).  For comparison,
we also show the $\rho^{\rm d}_1(a)$ distribution obtained by forcing
$\rho_2 (a, t) = 0$ while numerically solving
Eqs.\,\ref{EQ:RHO0}--\ref{kernels}.  Since this constraint allows only
non-radicals and activists to arise, $\rho^{\rm d}_1(a)$ may be used
as a proxy for the age distribution of activists at the dormant state;
we can also interpret $\rho^{\rm d}_1(a)$ as the age-dependent
analogue of the dormant solution discussed for the age-independent
model in subsection \ref{AI}. As can be seen from
Fig.\,\ref{FIG:CP0_AGE_DIST}(a), the $\rho_1(a, t=4.2)$ and $\rho^{\rm
  d}_1(a)$ distributions are almost identical, suggesting that the
system transitions through the dormant state while evolving from
utopia to turmoil, similarly as to what observed in the
age-independent case, as discussed in Fig.\,\ref{FIG:NO_AGE1}(c).  As
mentioned above, the age distributions corresponding to
Fig.\,\ref{FIG:AGE_TIME}(a) at $t=100$ would look qualitatively
similar to $\rho_i(a, t=4.2)$ for $i=1,2$ in
Fig.\,\ref{FIG:CP0_AGE_DIST}(a).

The parameters used in Fig.\,\ref{FIG:AGE_TIME}(c) lead to oscillatory
behavior.  Since a long-term steady state cannot be identified we plot
the age distributions at times of interest. We begin with $t = 1.8$
when the activist population reaches its first maximum, corresponding
to the early peak in Fig.\,\ref{FIG:AGE_TIME}(c), and the radical
population has increased to appreciable values. In
Fig.\,\ref{FIG:CP0_AGE_DIST}(b) the age distributions for both
populations peak near ages $a \simeq \alpha_{\rm A} = \alpha_{\rm R} =
20$, with $\rho_1(a, t=1.8) \ll \rho_2(a,t=1.8)$ for all ages.  At
later times, as the number of radicals increases, the number of
activists decreases to almost extinction. In
Fig.\,\ref{FIG:CP0_AGE_DIST}(b) we also plot $\rho_2(a,t=10)$ and
$\rho_2(a,t=20)$. Their shape is almost identical and $\rho_2(a,t=20)
\simeq \rho_2(a+10, t=10)$ since within this time-frame radicals can
only age, but neither increase nor decrease, as the activist pool is
depleted, and radicalization is irreversible.  Eventually, at $a =
a_1$, radicals age out of the system, repopulating first the
non-radical, and subsequently the activist pool. A second oscillation
cycle begins.

In Fig.\,\ref{FIG:CP0_AGE_DIST}(c) we plot the age distributions
beginning at $t = 61.7$, when the activist fraction reaches the second
maximum shown in Fig.\,\ref{FIG:AGE_TIME}(c).  Here the radical
population is vanishingly small while the activist age-distribution is
very close to that of dormant state.  The two behaviors at $t=1.8$ and
$t=61.7$ are different, due to initial conditions. In particular, we
note that at the onset even the small, uniform, radical population
$\rho_2(a,t=0)= 0.001$ is able to funnel non-radicals towards full
fledged extremism, bypassing the dormant state.  A large number of
radicals quickly consolidates. This initial condition effect damps
away after a transient period, when the number of radicals declines
due to aging and the non-radical population is replenished. As the
cycle begins anew with almost no radicals, the system is able to reach
the dormant state, with a large activist population. Eventually the
latter radicalize, yielding the age distribution $\rho_2(a, t=70)$
depicted in Fig.\,\ref{FIG:CP0_AGE_DIST}(c).  The above dynamics show
that while at short times the initial radical population serves as a
conduit to early, large scale radicalization, it also causes the rapid
depletion of the activist pool, until recruits are no longer
available, halting further radicalization. On the other hand, in the
second cycle, with a vanishingly small initial radical population,
activists are better able to increase their ranks and in turn to
sustain a continuous transition towards the final radical stage.
Indeed, the total number of radicals in the second cycle from both
Fig.\,\ref{FIG:AGE_TIME}(c) and and Fig.\,\ref{FIG:CP0_AGE_DIST}(c) is
much larger than in the first cycle.  The early emergence of radicals
may thus be counterproductive.  Advection is still observed as
$\rho_2(a, t = 80) \simeq \rho_2(a+10,t=70)$.

\begin{figure*}[t]
  \begin{center}
      \includegraphics[width=5.2in]{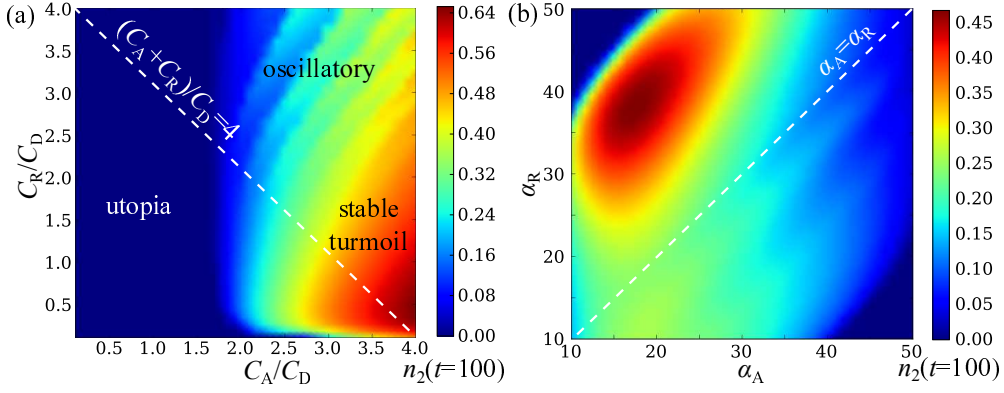}
\end{center}
  \caption{Long-time $n_2 (t = 100)$ radical population as a function
    of $C_{\rm A} / C_{\rm D}$ and $C_{\rm R} / C_{\rm D}$ in panel
    (a) and $\alpha_{\rm A}$ and $\alpha_{\rm R}$ in panel (b).  In
    panel (a) $n_2(t=100)$ is negligible until $C_{\rm A} / C_{\rm D}
    \gtrsim 2$, after which it increases. As a function of $C_{\rm
      R}/C_{\rm D}$, $n_2 (t=100)$ peaks at $C_{\rm R} / C_{\rm D}
    \simeq 0.5$; further increases lead to a decline and to
    oscillatory behavior. In panel (b) $n_2 (t=100)$ peaks at
    $\alpha_{\rm A} \simeq 15$ and $\alpha_{\rm R} \simeq
    35$. Starting from the maximum, $n_2 (t=100)$ drops sharply by
    increasing $\alpha_{\rm R}$ and gradually by increasing
    $\alpha_{\rm A}$. Unless being varied, parameters are $C_{\rm D} =
    5$, $C_{\rm A} = 12$, $C_{\rm R} = 8$, $\alpha_{\rm A} =
    \alpha_{\rm R} = 20$, and $\sigma_{\rm A} = \sigma_{\rm R} = 10$.
         \label{FIG:FRAC_T100} }
\end{figure*}

We now examine how other parameters influence the long-term radical
population. In Fig.\,\ref{FIG:FRAC_T100}(a) we plot $n_2 (t=100)$ as a
function of $C_{\rm A} / C_{\rm D}$ and $C_{\rm R} / C_{\rm D}$. In
general, $C_{\rm A} / C_{\rm D}$ is seen to dictate whether a radical
population may ever arise, while $C_{\rm R} / C_{\rm D}$ determines
long time behavior and whether a steady or an oscillatory state is
reached.  The $n_2 (t = 100)$ population increases with increasing
$C_{\rm A} / C_{\rm D}$ and reaches a maximum at $C_{\rm R} / C_{\rm
  D} \simeq 0.5$ regardless of $C_{\rm A} / C_{\rm D}$. As discussed
above, $n_2 (t = 100)$ has not reached steady state for smaller
$C_{\rm R}/C_{\rm D}$ and is hampered by over-drainage of the activist
pool for larger $C_{\rm R}/C_{\rm D}$. Finally for large $C_{\rm
  R}/C_{\rm D}$ oscillatory behavior arises, as indicated by the
striped pattern in the upper-right portion of
Fig.\,\ref{FIG:FRAC_T100}(a). Reducing $C_{\rm A}/ C_{\rm D}$ hinders
the activation of non-radicals and may be considered a prevention
strategy; in contrast, reducing $C_{\rm R}/C_{\rm D}$ impedes full
radicalization and may be considered a correctional strategy. The
results from Fig.\,\ref{FIG:FRAC_T100}(a) suggest that prevention may
be more effective than correction. To convey this point, we highlight
a line of constant $C_{\rm A}/C_{\rm D}+ C_{\rm R}/C_{\rm D} = 4$ to
represent a scenario where total escalation rates rates are constant
but can be preferentially directed towards activation or
radicalization. Under this constraint, lowest values of $n_2 (t=100)$
are found by decreasing $C_{\rm A}/C_{\rm D}$; although the
corresponding radicalization rate is large, the activist pool is too
small to yield a sizable radical population.
 
%
%
  
  Fig.\,\ref{FIG:FRAC_T100}(b) shows the steady state radical
  population as a function of $\alpha_{\rm A}$ and $\alpha_{\rm
    R}$. The maximum of $n_2 (t=100)$ is located at $ \alpha_{\rm A}
  \simeq 17$ and $\alpha_{\rm R} \simeq 38$, suggesting that
  radicalization is most pronounced when non-radicals are activated in
  their late teens and the resultant activist pool is given ample time
  to develop. The large age difference between $\alpha_{\rm A}$ and
  $\alpha_{\rm R}$ allows the age stratified activists to most
  effectively funnel non-radicals to the radical pool.  Although such
  large values of $\alpha_{\rm R}$ are not realistic
  Fig.\,\ref{FIG:FRAC_T100}(b) implies that, given an activation age
  $\alpha_{\rm A}$, the number of radicals increases with
  radicalization age $\alpha_{\rm R}$, allowing the activist pool to
  keep sustaining radical growth. This growth continues until
  $\alpha_{\rm R}$ reaches large values (here, roughly 40), when the
  various populations no longer overlap in age, and the interaction
  kernel ${\mathcal{K}}(a,a'; \alpha_{\rm R}, \sigma_{\rm R})$ loses
  its effectiveness.
   
So far we have analyzed expected outcomes over long times.  In
particular, we find that age-dependence allows for the asymptotic
emergence of utopia, a dormant state and turmoil, just as in the
age-independent formulation.  However, the basins of attraction of
these steady states and the associated populations may differ greatly
in the two formulations. For example, the strong promotion of
radicalization at early ages may lead to a premature draining of the
activist pool, effectively thwarting further radicalization.  In this
context, the radical ideology spreads too quickly among a few who
become isolated from the rest of society and who are not able to
effectively recruit more adherents through peer-pressure at the
intermediate, activist stage.  Most notably, in certain parameter
regimes, age-structure allows for cyclic behavior to arise, with
alternating waves of more or less radicalized individuals. The above
described trends often arise over several years, when the system has
equilibrated.  In more practical situations, however, interventions
may have been introduced earlier, before reaching the above steady
state predictions.  We thus turn our attention to short-time
phenomena, at the onset of an escalating situation.

\subsection{Short-time behavior}

\noindent
As outlined in the previous subsection, starting from a quasi-utopia
initial configuration, the dynamics of the system evolves over two
stages. Activists emerge first and reach a metastable, finite
population; radicals remain few. In the second phase, activists turn
into radicals, and the metastable state is dissipated.  In this
subsection we use perturbation theory to examine the onset of the
activist population from an initial non-radical society and the onset
of the radical population from the activist pool.

We begin by perturbing utopia, defined as $\rho_i(a) = 1$ for $i=0$
and $\rho_i(a) = 0$ for $i=1,2$ and examine how trajectories depart
from it. An age-dependent perturbation around utopia can be expressed
as $\rho_1(a, t) = \delta \rho_1 (a, t)$, $\rho_2 (a, t) = \delta
\rho_2 (a, t)$, and $\rho_0(a, t) = 1 - \delta \rho_1 (a, t) - \delta
\rho_2 (a, t)$, where $\delta \rho_1 (a, t), \delta \rho_2 (a, t) \ll
1$ are small perturbations.  The age structures at $t=0$ are given as
$\delta \rho_i(a,t=0)$ for $i=1,2$.  Upon linearizing the system to
first order we find that for short times

%
\begin{figure*}[t]
  \begin{center}
      \includegraphics[width=5.15in]{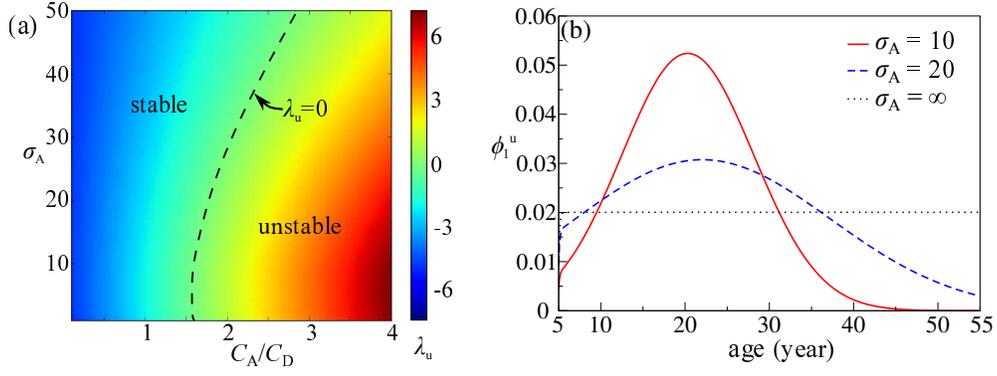}
\end{center}
  \caption{ Perturbations around utopia.  (a) Activist growth rate
    $\lambda_{\rm u}$ as a function of $C_{\rm A} / C_{\rm D}$ and
    $\sigma_{\rm A}$.  The eigenvalue $\lambda_{\rm u}$ decreases with
    increasing $C_{\rm A} / C_{\rm D}$ and decreasing $\sigma_{\rm
      A}$, suggesting that kernels tightly clustered in age magnify
    the instability of utopia.  (b) The age-differentiated growth
    amplitude $\phi^{\rm u}_1(a)$, the eigenfunction corresponding to
    $\lambda_{\rm u}$, for various $\sigma_{\rm A}$ with $C_{\rm A} /
    C_{\rm D} = 2.4$ and $\alpha_{\rm A} = 20$.  As $\sigma_{\rm A}$
    increases, the peak of $\phi_1^{u}(a)$ shifts towards $a >
    \alpha_{\rm A}$ and its width increases as described in the
    text. For even larger $\sigma_{\rm A}$, $\phi_1^{u}(a)$ is
    quasi-constant with the exception of the region $a \simeq a_0$,
    where the boundary condition suppresses activist growth by
    enhancing the non-radical pool.  Unless being varied, parameters
    are $C_{\rm A} = 12$, $C_{\rm D} = 5$, $C_{\rm R} = 5$,
    $\alpha_{\rm A} = 20$.
     \label{FIG:EVALUE_UTOPIA}}
\end{figure*}
%

\begin{eqnarray}
   \frac{\partial \delta \rho_1} {\partial t} + \frac{\partial \delta \rho_1 }{\partial a} 
   &=& A (a; \delta \rho_1) - C_{\rm D} \delta \rho_1, 
    \label{EQ:DELTARHO1A}  \\
  \frac{\partial \delta \rho_2 }{\partial t} + \frac{\partial \delta \rho_2 }{\partial a} 
  &=&  0,
   \label{EQ:DELTARHO2A} 
\end{eqnarray}

\noindent 
where $A(a; \delta \rho_1)$ is given by
Eq.\,\ref{EQ:ACTIVATION_RATE}. The perturbations do not contribute to
the boundary condition in Eq.\,\ref{EQ:BIRTH1} since their sum is zero
by construction.  Eq.\,\ref{EQ:DELTARHO2A} implies that $\delta
\rho_2(a,t)$ retains its original age structure and is simply advected
in time.  The radical population remains small for short times as
expected.  Since it is linear in $\delta \rho_1$,
Eq.\,\ref{EQ:DELTARHO1A} can be recast in terms of an age-dependent
operator ${\cal L}(a)$ defined as

\begin{eqnarray}
  \frac{\partial \delta \rho_1} {\partial t} = {\cal L}(a) \delta \rho_1.
    \label{EQ:DELTARHO3A} 
\end{eqnarray}

\noindent
The growth of an initially infinitesimally small $\delta \rho_1(a,t)$
can be approximated as $\delta \rho_1 (t, a) \sim \exp(\lambda_{\rm u}
t) \phi_1^{\rm u} (a)$, where $\lambda_{\rm u}$ is the largest
eigenvalue arising from Eqs.\,\ref{EQ:DELTARHO3A} and $\phi_1^{\rm u}
(a)$ is its corresponding eigenfunction.  We can interpret
$\lambda_{\rm u}$ as the onset growth rate of activists: $\lambda_{\rm
  u} < 0 $ implies a regression of activists towards utopia and
characterizes the perturbations as unstable, whereas $\lambda_{\rm u}
> 0$ indicates departure from utopia and characterizes the
perturbations as stable.  Unless otherwise stated or varied,
parameters are chosen here as $C_{\rm A} = 12$, $C_{\rm D} = 5$,
$C_{\rm R} = 2$, $\sigma_A = \sigma_R = 10$, $\alpha_A = \alpha_R =
20$, as used in Fig.\,\ref{FIG:AGE_TIME}(b).  We numerically compute
the largest eigenvalue $\lambda_{\rm u}$ and its corresponding
eigenfunction $\phi_1^{u}(a)$ using the power method
\cite{MIS29,BOD56,GOL00}.  Here, the given operator is applied to an
arbitrary initial function, the result is normalized and the result,
and iteratively multiplying the operator to the normalized result
until the normalized result converges.

In Fig.\,\ref{FIG:EVALUE_UTOPIA}(a) we plot $\lambda_{\rm u}$ as a
function of $C_{\rm A} / C_{\rm D}$ and $\sigma_A$, the relative
intensity and width of the activation rate. For low values of $C_{\rm
  A} / C_{\rm D}$, $\lambda_{\rm u}$ is negative: perturbations vanish
and utopia is stable. Higher values of $C_{\rm A} / C_{\rm D}$ lead to
positive $\lambda_{\rm u}$ and to large scale instabilities. The
marginal stability line at $\lambda_{\rm u} = 0$ separates the two
regimes. The unstable region expands with decreasing $\sigma_{\rm A}$,
suggesting that non-zero activist configurations may be more easily
reached by narrowing the age range in the interaction kernel ${\cal K
} (a,a'; \alpha_{\rm A}, \sigma_{\rm A})$, while at the same time
increasing its peak by construction.  Larger $\sigma_{\rm A}$ values
spread the kernel over a large age range, resulting in a modest peak
that may not be enough to sustain activist growth.  As expected,
$\lambda_{\rm u}$ increases with $C_{\rm A} / C_{\rm D}$.  Other
parameters $\alpha_{\rm A}$, $C_{\rm R}$, $\sigma_{\rm R}$ and
$\alpha_{\rm R}$ have negligible effects.

The eigenfunction $\phi_1^{\rm u} (a)$ represents the
age-differentiated growth amplitude of activists and is shown in
Fig.\,\ref{FIG:EVALUE_UTOPIA}(b) for three $\sigma_{\rm A}$ values.
As is evident for increasing $\alpha_A$, $\phi_1^{\rm u}(a)$ peaks at
$a \gtrsim \alpha_{\rm A}$, not at $a = \alpha_A$. This can be
explained by recalling that activist populations increase with age,
since we initiate our system from a quasi-utopia state, and since our
chosen boundary conditions in Eqs.\,\ref{EQ:BIRTH1} and
\ref{EQ:BIRTH2} preferentially repopulate the non radical state at age
$a=a_0$.  Since the interaction kernel ${\cal K}(a,a'; \alpha_A,
\sigma_A)$ favors interactions between same-age individuals,
non-radical early adults, slightly older than $a = \alpha_{\rm A}$,
are more likely to interact with activists of their same age than with
non-radicals, as activists are simply more numerous at this age.
Hence, there is a higher likelihood that young non-radical adults
older than $a = \alpha_{\rm A}$ become activated due to societal
interactions.  The peak of $\phi_1^{u}(a)$ is shifted further to the
right as the range of social interactions increases, that is, for
larger values of $\sigma_{\rm A}$.  Of course, this outcome depends on
the specific choice for ${\cal K}(a,a'; \alpha_A, \sigma_A)$.
Finally, as can be seen, $\phi_1^{\rm u} (a)$ becomes wider and
shallower with increasing $\sigma_{\rm A}$, until, for $\sigma_{\rm A}
\to \infty$, $\phi_1^{\rm u}(a \gg a_0)$ is constant, except for
$\phi_1^{\rm u}(a \simeq a_0)$, where boundary conditions pin it close
to zero.

In the unstable case, when $\lambda_{\rm u} >0$, perturbations around
utopia grow and drive the system towards the dormant steady state.
Note that $\phi_1^{\rm u}(a)$ shown in
Fig.\,\ref{FIG:EVALUE_UTOPIA}(b) for $\sigma_{\rm A} = 10$ and
$\rho_1^{\rm d}(a)$ shown in Fig.\,\ref{FIG:CP0_AGE_DIST}(c) have
similar profiles, suggesting that as the system evolves from utopia to
the dormant state, the $\phi_1^{\rm u} (a)$ eigenfunctions grow
uniformly towards $\rho_1^{\rm d} (a)$.

At longer times, the system leaves the dormant region and progresses
towards the turmoil state.  We similarly investigate the short-term
behavior of this process using perturbation analysis. In particular,
we reset time so that at $t=0$ the system is in the dormant state and
model small fluctuations as $\rho_1 (t, a) = \rho_1^{\rm d} (a) +
\delta \rho_1 (t, a)$, $\rho_2 (t, a) = \delta \rho_2 (t, a)$, and
$\rho_0 (t, a) = 1 - \rho_1^{\rm d} (a) - \delta \rho_1 (t, a) -
\delta \rho_2 (t, a)$, where the perturbations $\delta \rho_1 (a,t),
\delta \rho_2 (a,t) \ll 1$ for all ages and for short times. Upon
substitution into Eqs.\,\ref{EQ:RHO0}-\ref{EQ:RHO2}, and linearizing
to first order we find

\begin{eqnarray}
\label{EQ:DELTARHO1B}
   \frac{\partial \delta \rho_1} {\partial t} + \frac{\partial \delta \rho_1 }{\partial a} 
   &=& - A (a; \rho_1^{\rm d}) \left( \delta \rho_1 + \delta \rho_2 \right)  - C_{\rm D} \delta \rho_1
   \\
   \nonumber
   && + A (a; \delta \rho_1) \left( 1 - \rho_1^{\rm d} \right) 
    - R(a; \delta \rho_2) \rho_1^{\rm d},  
\\
  \frac{\partial \delta \rho_2 }{\partial t} + \frac{\partial \delta \rho_2 }{\partial a} &=&  
   \label{EQ:DELTARHO2B} 
   R (a; \delta \rho_2) \rho_1^{\rm d}.
\end{eqnarray}

\noindent
As done above, we also approximate the growth of the perturbation near
the dormant state as $\delta_i (t, a) \simeq \exp( \lambda_{\rm d} t )
\phi_i^{\rm d} (a)$, where $\lambda_{\rm d}$ is the largest eigenvalue
of the operator stemming from Eqs.\,\ref{EQ:DELTARHO1B} and
\ref{EQ:DELTARHO2B} and represents the growth rate of radicals near
the dormant state.  The corresponding eigenfunctions are $\phi_i^{\rm
  d}(a)$ with $i = 1, 2$.

%
\begin{figure*}[t]
  \begin{center}
      \includegraphics[width=6.9in]{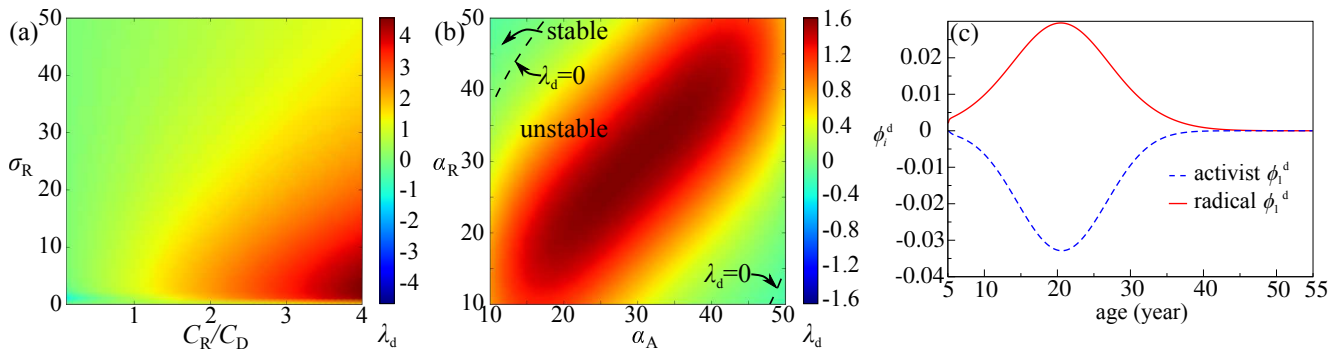}
\end{center}
  \caption{ Perturbations around the dormant state.  (a) Radical
    growth rate $\lambda_{\rm d}$ as a function of $C_{\rm R} / C_{\rm
      D}$ and $\sigma_{\rm R}$.  The dormant state is mostly unstable
    since $\lambda_{\rm d}$ is always positive.  (b) Radical growth
    rate $\lambda_{\rm d}$ as a function of $\alpha_{\rm A}$ and
    $\alpha_{\rm R}$.  $\lambda_{\rm d}$ is greatest near the
    $\alpha_{\rm A} = \alpha_{\rm R}$ line. Small patches of stable
    regimes occur when $\alpha_{\rm R}$ and $\alpha_{\rm A}$ differ
    greatly.  (c) Normalized growth patterns of activists $\phi_1
    (a)$, and radicals, $\phi_2 (a)$. Radicals increase at the expense
    of activists, since $\phi_2(a) > 0$ and $\phi_1(a) < 0$.  Unless
    varied, parameters are $C_{\rm A} = 12$, $C_{\rm R}$ = 8,
    $\alpha_{\rm A,R} = 20$, $\sigma_{\rm A,R} = 10$, $C_{\rm D} = 5$,
    $\sigma_{\rm D} \to \infty$.
         \label{FIG:EVALUE_DORMANT}}
\end{figure*}
%

In Fig.\,\ref{FIG:EVALUE_DORMANT}(a) we plot $\lambda_{\rm d}$ as a
function of ${C_{\rm R}}/{C_{\rm D}}$ and $\sigma_A$. As can be seen,
the dormant state is always unstable: in the absence of a pacification
mechanism, radicals will never return to the activist stage. In
Fig.\,\ref{FIG:EVALUE_DORMANT}(b) we analyze the dependence of
$\lambda_{\rm d}$ on $\alpha_{\rm R}$ and $\alpha_{\rm A}$.  We find
that $\lambda_{\rm d}$ reaches a maximum for $\alpha_{\rm R} \gtrsim
\alpha_{\rm A}$, at approximately the same age when the activist
distribution in the dormant state $\phi_1^{\rm d}(a)$ also reaches a
maximum. Thus, radicalization may be greatly enhanced by targeting the
ages corresponding to the largest activist population.  It is
interesting to note that the long-term behavior of $n_2(t=100)$ and
the short-term radical growth rate $\lambda_d$ depend on $\alpha_{\rm
  A}$ and $\alpha_{\rm R}$ in very different ways, as can be seen by
comparing Figs.\,\ref{FIG:FRAC_T100}(b) and
\ref{FIG:EVALUE_DORMANT}(b).  At short times,
Fig.\,\ref{FIG:EVALUE_DORMANT}(b) shows that $\lambda_d$ is largest
for $\alpha_{\rm R} \simeq \alpha_{\rm A}$; at longer times,
\ref{FIG:FRAC_T100}(b) reveals that $n_2(t=100)$ is largest for
$\alpha_{\rm R} > \alpha_{\rm A}$.  These results indicate that
although the instability is initially largest when the two target ages
are similar, the choice of $\alpha_{\rm R} \simeq \alpha_{\rm A}$
drains the activist pool too quickly for radicals to achieve long term
growth.  Radical populations are largest when the two target ages are
different allowing for the age-structured interactions to optimally
funnel populations to the radical state.

In Fig.\,\ref{FIG:EVALUE_DORMANT}(d) we compute the eigenfunctions
$\phi_1^{\rm d}(a)$ and $\phi_2^{\rm d} (a)$ for $C_{\rm R}/C_{\rm D}
= 0.4$, with the other parameters as detailed above.  These
eigenfunctions represent the age-differentiated growth amplitudes of
activists and radicals, respectively.  Numerically, we find
$\lambda_{\rm d} = 0.37$.  A positive $\phi_2^{\rm d} (a)$ and a
negative $\phi_1^{\rm d}(a)$ emerge, indicating growing radical and
decreasing activist populations, as expected. However, $\phi_2^{\rm
  d}(a)$ does not exactly match $-\phi_1^{\rm d}(a)$: the latter
exceeds the former at young ages, implying that a fraction of
activists returns to the non-radical status. At old ages $\phi_2^{\rm
  d}(a)$ exceeds $-\phi^{\rm d}_1(a)$, suggesting that non-radicals
are activated, and subsequently radicalized, spurring further
radicalization.  As the system evolves away from utopia or the dormant
state, a governmental agency may intervene to stem radicalization
through age-based educational or rehabilitation programs.  In the next
section, we explore the effects of age-dependent intervention.

\section{Government intervention}
\label{govt}

\noindent
To study the effects of government sponsored programs, we consider a
population that undergoes radicalization from an initial utopia and
irreversibly. We model intervention in the form of age-dependent
conversion terms that lower extremist levels incrementally, converting
radicals to activists, or activists to non-radicals. We thus re-write
the previously uniform de-escalating terms $C_{\rm D}$ and $C_{\rm
  P}=0$ in Eqs.\,\ref{EQ:RHO0}--\ref{EQ:RHO2}

\begin{eqnarray}
& & C_{\rm D} \to C_{\rm D} + \mu G (a), 
\label{CDage}\\
& & C_{\rm P} =0 \to (1 - \mu) G (a),
\label{CPage}
\end{eqnarray}

\noindent
where $0 \le \mu \le 1$ determines how efforts are divided between the
two possible intervention avenues.  Relatively large $\mu$ implies
that most resources are focused on returning activists to the
non-radical state, while more modest $\mu$ values represent
intervention in returning radicals to the activist state.  The age
dependent term $G(a)$ is modeled similarly to the interaction kernels
in Eq.\,\ref{kernels}

\begin{equation}
  G (a) = C_{\rm G} \frac{\displaystyle{\exp \left[ - \frac{(\alpha_{\rm G} - a)^2}{2 \sigma_{\rm G}^2} \right]}}
  {\displaystyle{\int_{-\bar{a}}^{\bar{a}} \exp \left[ -\frac{s^2}{2 \sigma_{\rm G}^2} \right] \dd s}}.
   \label{EQ:GOV}
\end{equation}

\noindent
Here, $C_{\rm G}$ represents the intensity of intervention, and
$\alpha_{\rm G}$ and $\sigma_{\rm G}$ are the corresponding target age
and spread.  To quantify the effectiveness of intervention, we
introduce a radical suppression ratio $\Psi \equiv {n_2^{\rm G}}(t \to
\infty)/ {n_2^{\rm T}(t \to \infty)}$ where $n_2^{\rm T}(t \to
\infty)$ is the steady state radical fraction at turmoil arising from
Eqs.\,\ref{EQ:RHO0}--\ref{kernels} and $n_2^{\rm G}(t \to \infty)$ is
similarly evaluated except for the government intervention
substitutions in Eqs.\,\ref{CDage} and \ref{CPage}.  To be concrete,
we select societal parameters as in Figs.\,\ref{FIG:AGE_TIME} and
\ref{FIG:AGE_TIME2}, $C_{\rm A} = 12$, $C_{\rm D} = 5$, $\alpha_{\rm
  A} = \alpha_{\rm R} = 20$, $\sigma_{\rm A} = \sigma_{\rm_R} = 10$,
initial conditions $\rho_0(a, 0) = 0.989$, $\rho_1(a, 0) = 0.01$,
$\rho_2(a, 0) = 0.001$, and vary $C_{\rm R}$.  Since $\Psi$ is
meaningful only if $n_2^{\rm G}(t \to \infty)$ and $n_2^{\rm T}(t \to
\infty)$ converge to unique values, we limit the $C_{\rm R} \le 10$,
before the oscillatory regime shown in Fig.\,\ref{FIG:AGE_TIME2} is
reached.  Government intervention is introduced once society has
reached steady state.

\begin{figure*}[t]
  \begin{center}
      \includegraphics[width=6.9in]{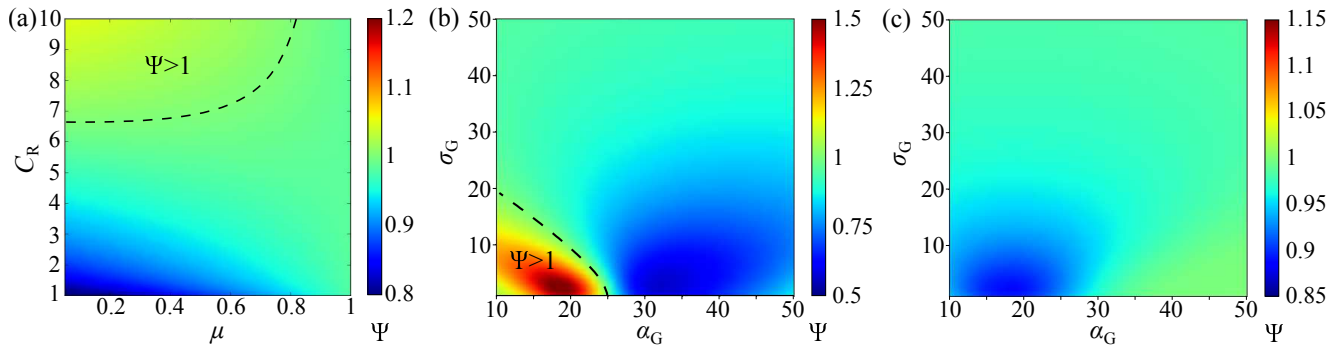}
\end{center}
  \caption{Government intervention.  (a) Radical suppression ratio
    $\Psi$ as a function of $\mu$ and $C_{\rm R}$ for $\alpha_{\rm G}
    = \alpha_{\rm A,R} $ and $\sigma_{\rm G} = \sigma_{\rm A,R}$. For
    strong radicalization (large $C_{\rm R}$) $\Psi$ decreases when
    $\mu$ increases, suggesting that activist prevention is more
    effective. For mild radicalization (low $C_{\rm R}$), radical
    prevention is more advantageous.  (b) Radical suppression ratio
    $\Psi$ as a function of $\alpha_{\rm G}$ and $\sigma_{\rm G}$
    through the pacification of radicals, $\mu = 0$. Optimal reduction
    of radicals is achieved for $\alpha_{\rm G} > \alpha_{\rm A,R} =
    20$.  Intervention may be counterproductive for targeted ages
    $\alpha_{\rm G} < \alpha_{\rm A,R} = 20$.  (c) Radical suppression
    ratio $\Psi$ as a function of $\alpha_{\rm G}$ and $\sigma_{\rm
      G}$ through the de-escalation of activists, $\mu = 1$. Optimal
    reduction of radicals is achieved for $\alpha_{\rm G} <
    \alpha_{\rm A,R} = 20$.  Unless varied, parameter values are
    $C_{\rm A} = 12$, $C_{\rm R} = 10$, $C_{\rm D} = 5$, $\alpha_{\rm
      A,R}= 20$, $\sigma_{\rm A,R} = 10$, and $C_{\rm G} = 1$ for (a)
    and $5$ for (b) and (c).
         \label{FIG:GOV_PARAMETER}}
\end{figure*}

As expected, $\Psi$ decreases with $C_{\rm G}$, indicating that larger
intervention leads to the emergence of less radicals.  In
Fig.\,\ref{FIG:GOV_PARAMETER}(a) we fix $C_{\rm G} = 1$, $\alpha_{\rm
  G} = \alpha_{\rm R} = \alpha_{\rm A}$, $\sigma_{\rm G} = \sigma_{\rm
  R} = \sigma_{\rm A}$, and plot $\Psi$ as a function of $\mu$ and
$C_{\rm R}$.  For small values of $C_{\rm R}$, $\Psi$ decreases with
$\mu$; the trend is reversed for large $C_{\rm R}$.  This suggests
that if the push towards radicalization is mild, at low $C_{\rm R}$,
the most effective intervention is to aim resources at pacification,
returning radicals to the activist state. Decreasing the number of
radicals will lessen their influence on potential recruits.
Vice-versa, for strong radicalization under large $C_{\rm R}$, the
optimal strategy is to encourage activists to return to the
non-radical stage, preventing radicalization at its incipit.  We also
note that $\Psi$ increases with $C_{\rm R}$, indicating that the more
aggressive radicalization is, the harder it is to suppress radicals.
Finally, when $C_{\rm R}$ is large and $\mu \to 0$, $\Psi$ can
increase beyond unity: government intervention enhances
radicalization. The mechanism behind this unwanted outcome is that
fostering the return of radicals towards the activist state yields a
much enlarged activist pool that is able to recruit more non-radicals
into the process. The net result is that the number of radicals will
increase. This effect is magnified for large $C_{\rm R}$, when
activists quickly turn into radicals, and the activist pool becomes a
scarcely populated bottleneck.  Government intervention enlarges the
activist pool, alleviating the bottleneck, and fortuitously aiding the
conversion of radicals.  We do not observe the same result at small
$C_{\rm R}$ since there is no bottleneck and activists are quite
numerous even before intervention.
%
%

  In Figs.\,\ref{FIG:GOV_PARAMETER}(b) and (c) we set $C_{\rm R} = 10$
  and $C_{\rm G} = 5$ and plot $\Psi$ as a function of the
  characteristic ages $\alpha_{\rm G}$ and $\sigma_{\rm G}$ for $\mu =
  0$ in panel (b) and $\mu = 1$ in panel (c).  Only values of
  $\sigma_{\rm G} \lesssim 10$ affect $\Psi$, which may be expected
  since the age-dependent model converges to the age-independent one
  as $\sigma_{\rm G} \to \infty$. In Fig.\,\ref{FIG:GOV_PARAMETER}(b),
  where $\mu=0$, $\Psi > 1$ for $15 \lesssim \alpha_{\rm G} \lesssim
  25$ and $\alpha_{\rm A} = 20$, implying government intervention is
  counterproductive, as outlined above. Vice-versa, $\Psi$ is lowest
  for $30 \lesssim \alpha_{\rm G} \lesssim 45$ when radicals who
  return to the activist pool interact less with younger non-radicals.
  In Fig.\,\ref{FIG:GOV_PARAMETER}(c), where $\mu = 1$, $\Psi < 1$ for
  all values of $\alpha_G$ and $\sigma_G$ and its lowest value is
  within $15 \lesssim \alpha_{\rm G} \lesssim 25$.

\section{Discussion}

\noindent
Several population dynamics models have been proposed to study
multi-stage radicalization \cite{CAS03,SAN08,MAR12,CAM13,EHR13,DEU15}.
Our work shows that including age dependence, even in the simplest
form of interactions between populations, can lead to rich dynamics,
enhancing radicalization in certain parameter regimes, and leading to
oscillatory behavior.  Age-independent radicalization models generally
converge to fixed ratios among populations and phase portraits do not
display limit cycles \cite{CAS03,DEU15}, as shown here for the
age-independent formulation of our model. Yet, according to one of the
most influential theories of terrorism, developed by the political
scientist D. C. Rapoport, extremist tendencies rise and fall over time
like waves or ripples \cite{RAP02,SED07,MCA11,PAR16}.  Our age structured
model provides a potential mechanism to explain such wave-like
behavior. In particular, we predict oscillatory solutions when i)
radicalization is aggressive ($C_{\rm R}$ is large), and ii)
radicalization is irreversible ($C_{\rm P} = 0$). As shown
  in Fig.\,\ref{FIG:AGE_TIME}(c), the typical timescale for sustained
  radical populations is roughly 40 years, in agreement
with sociological observations \cite{RAP02,SED07,RAS09,MCA11,PAR16}.
The lifetime of
these radical states
depends on the age interval $[a_0, a_1]$
within which existing radicals are able
to convert non-radicals.
This finding
supports Rapoport's conjecture that extremists 
radicalize a generation of individuals, and when their
influence fades due to aging, the cycle of terrorism comes to an 
end \cite{RAP02,MCA11}.
Recent 
studies using data analysis to quantitatively inspect Rapoport's wave
theory have observed a relatively shorter lifespan of about $30$ years for the so-called
``New Left'' terrorist wave emerging from Marxist
movements \cite{RAS09}. This observation is consistent with a smaller
effective age interval of influence in our model, 
and can be justified by noting 
that Marxism is a relatively more sophisticated ideology,
compared to religious or nationalist fanaticism. 
Indoctrination of younger individuals may thus be less effective,
since their understanding of socioeconomic issues may still be under
development.  
Quantitative studies have been facilitated by the creation of
dedicated databases to index terrorist related data starting from the
1960s \cite{BOG05,LAF07}. Prior records are sparse and
the data is not sufficient for quantitative analysis of
radical resurgence within a society and over long periods of time, as would be
required to validate the existence of generational waves predicted by
our model.  
Global data, such as fatalities in conflicts over the past several
centuries do exhibit oscillatory patterns with a period of roughly
half a century \cite{ROS}; however lumping together data from various
parts of the world may not be directly applicable to our model, due to
the great regional variability and the limited information spread of
centuries past. 
From a more qualitative perspective, religious driven conflicts seem
to display signs of ebbs and flows throughout history at regional
scales.
For examples, 
the 13th century Crusades in Europe were followed by the Renaissance,
a relatively peaceful era, and again by the 16th century protestant
Reformation wars. Similarly, in the Middle East,
the 19th century Islamic revival inspired anti-colonial wars; 
religious leaders were later replaced by secular authorities who ruled
for most of the 20th century,
until they were toppled or threatened by religious movements. 
Notably the worldwide decline of religious
influence in the 20th century even led to the once prominent
``secularization'' theory \cite{CHA94,WAL06,FOX12}, according to which
religions would  
continue fading and eventually disappear. However, religious
fanaticism has returned with a vengeance in recent 
decades \cite{WAL06,FOX12}.
Oscillatory phenomena can sometimes be observed within
long-lasting conflicts, such as the struggle between Israelis and
Palestinians, although counter-measures implemented in response to 
specific incidents can contribute to some short-term fluctuations as
well \cite{ISR}.
Admittedly the real world is much more complicated than our model, and the
evolution of human society will involve other factors that
are not included in Eqs.\,\ref{EQ:RHO0}--\ref{kernels}.
Notwithstanding, the oscillatory solutions of our model under fixed
conversion intensities $C_{\rm A/R/D}$ suggest the
possibility of resurgence for a declining or even inactive ideological
movement, provided that the underlying sociological contexts 
persist over time.

In addition to cyclic behavior, we also find that age-structure
increases the sensitivity of the dynamics to parameter values. For
example, in Fig.\,\ref{FIG:AGE_TIME2} the long-time radical fraction
$n_2 (t=100)$ derived in the age-dependent model is nearly doubled by
doubling $C_{\rm R}$ from $1$ to $2$, while the corresponding
age-independent $n_2 (t=100)$ stays uniform.  Moreover, as can be seen
in Fig.\,\ref{FIG:FRAC_T100}(b) the age-dependent $n_2 (t =100)$ is
highly sensitive to shifts in $\alpha_{\rm A}$ and $\alpha_{\rm R}$,
even if the magnitudes of the rates stay unchanged. These results
indicate that age-dependent radicalization models can lead to more
volatility and to more complex behavior compared to the more
straightforward predictions arising from age-independent models. These
nuances imply that governments can greatly improve the outcome of
their interventions by targeting the right age groups, rather than
treating all ages the same way.

Finally, we find that the most effective tactic to reduce the radical
population is to prevent activation by decreasing $C_{\rm A}$ as shown
in Fig.\,\ref{FIG:FRAC_T100}(a).  From a practical perspective, this
may be achieved by reducing the factors that lead to the activation of
non-radicals through education, hope for the future, employment.
Sometimes however, this may not be possible and eliminating escalating
factors for one group may lead another to radicalize. An alternate,
corrective approach, may be to de-activate the activist population or
to pacify radicals.

In cases where radicalization is very aggressive, the optimal
intervention strategy is to reduce the number of activists,
effectively isolating the most extreme from the general non-radical
public. However, when radicalization is moderate, although the number
of radicals is relatively small, a large body of activists will
emerge, in response to an ideology that may be justified among the
general public.  In this case, the best intervention strategy is to
focus on pacification, reducing the numbers of those who have taken
the ideology to extreme levels.  Current intervention policies are of
two types: stopping extremists who are about to engage in violent
activities, while allowing all freedom of speech is known as the
``Anglo-Saxon'' approach; preventing the spread of extreme views
before violence arises is the ``European'' approach \cite{NEU13a}.  We
can loosely interpret the choice of $\mu=0$ in our intervention
protocol to represent the Anglo-Saxon approach, where radicals are
targeted, while $\mu=1$ corresponds to the European approach, where
the de-escalation of activists is sought.  Our results suggest that
discerning which strategy is best may depend on the aggressiveness of
the radicalization process, embodied by $C_{\rm R}$.

\section{Conclusion}

\noindent
In this paper we introduced a multi-stage radicalization model with
age-structured progression rates. Upon comparison with the
corresponding age-independent formulation, we find that age dependence
leads to more complex, parameter-sensitive dynamics.  In many cases,
especially for irreversible radicalization, age dependence enhances
the number of irreducible radicals. For large radicalization rates,
age structure leads to oscillatory behavior, where large fractions of
extremists ebb and flow over a lifetime of roughly $40$ to $50$ years.
While the enhanced parameter sensitivity implies higher risks for
escalation, it also provides an opportunity for effective policy
making, for example by surgically targeting more susceptible age
groups. Upon comparing different government intervention strategies,
we find that de-activation, i.e. returning mildly indoctrinated
individuals to the non-radical state, leads to more suppression of
extremists when the radicalization rate is high. In the opposite case,
the pacification of radicals may produce better results.
Despite the great simplicity of the mathematical model, 
we observe a wide range of dynamical behaviors simply by distributing the
strength of social interactions differently over age, without
including more complicated external factors, such as the 
evolution of socioeconomic contexts or the change of political regimes. 
Sociologically our findings suggest that the heterogeneity of social
interactions among different age groups can profoundly change the course of 
progression, e.g., the longevity and the
pinnacle point, of an ideology within a population. Fundamentally
distinctive phenomena can be overlooked if the effect of age
structure is neglected and all age groups are lumped into a uniform population. 
It is particularly interesting to note that the lifespan of radical
activities in our oscillatory scenarios quantitatively agree with the
observed time scale in the wave theory of terrorism \cite{RAP02,RAS09,MCA11}. Our model further
suggests the potential for a past terrorist wave to resurge even after a long period
of dormancy, unless the sociological contexts underlying the
previous wave is sufficiently reduced to ensure the stability of 
a peaceful utopia. 

Our model includes only three linear stages of radicalization. In
particular, all non-violent steps are condensed in the intermediate
activist stage, between the general non-radical population and the
violence-prone extremists.  We may increase the number of stages as
done in previous age-independent work
\cite{BOR03,WIK04,MOG05,PRE07,SAG08,KLA16} for a more nuanced
progression to radicalization, but we expect qualitatively similar
results to what showed here for a linear pathway. Our model may be
best used to approximate a slowly evolving society, where the
populations are given sufficient time to relax toward quasistatic
conditions, allowing higher order interactions, say, between $\rho_0$
and $\rho_2$, to dampen away, and a unique, primary pathway to emerge.
Our findings can be very different from those arising in more
dynamical situations where multiple pathways exist to radicalization,
or more than two stages are nonlinearly coupled in an interaction.
For example, social interactions may exhibit ``history'' dependence,
where the likelihood of radicalization may depend on the entire
sequence of past states. Such history dependence may be directly
incorporated in the transition rates between radicalization stages
\cite{CHU17}, or through the notion of social reinforcement, where
transitions require multiple stimulations \cite{CEN10}.

Furthermore, we assumed a constant total population with a uniform age
distribution, and have not included birth and death events
\cite{CHO16}.  Also not included is the possibility of forcefully
removing individuals from the radical or activist pools, via arrest or
involuntary exile.  Neglecting birth and death may be a more suitable
assumption for developed countries \cite{HOW11}, but for developing
countries, especially in war-torn regions, birth and death rates are
high and not balanced. Birth and death are known to have profound
impact on the age structure of the overall population, leading to
non-uniform age distributions \cite{BOC94,KEY97,CHO16}.  Previous
studies have connected the development of civil conflicts with a
particular type of non-uniform age distributions known as the ``youth
bulge'' where the population consists of disproportionately large
youth cohorts \cite{FUL90,HAR04,HAR05}. It is believed that youth
bulge may intensify the competition for resource and employment
opportunities, further exacerbating feelings of disaffection among the
young \cite{MES99,URD06,YAI16,FAR17}.  The effect of age-dependent
birth and death have been studied under the same McKendrick-von
Foerster framework presented here for various biological and
ecological applications
\cite{MCK26,LES45,LES48,TRU65,BOC94,KEY97,CHO16} and can be added as
an extension of our model to investigate youth bulge or related
phenomena. While incorporating birth and death is beyond the scope of
this paper, our current study can be used as a baseline for
age-dependent social interactions, upon which further complexities may
be added to describe more realistic scenarios.

Lastly we present only the long-term outcome of government
intervention policies, discussing what strategies may achieve the best
results, regardless of implementation cost.  Given limited resources,
the best strategy may not always be possible, so a utility function
should be derived to seek an optimal balance between the cost of
intervention and the gain from the prevention of radicalization.  For
example, a time-dependent utility function may allow the government to
dynamically adjust its strategy based on past outcomes
\cite{SHO13,SHO17}.


\bibliography{crime}




\end{document}